\documentclass[manuscript,screen]{acmart}

\usepackage{graphicx}
\usepackage{booktabs}
\usepackage{multirow}
\usepackage{caption}
\usepackage{subcaption}
\usepackage{color}
\usepackage[linesnumbered,ruled,vlined]{algorithm2e}

\AtBeginDocument{%
  \providecommand\BibTeX{{%
    \normalfont B\kern-0.5em{\scshape i\kern-0.25em b}\kern-0.8em\TeX}}}

\setcopyright{acmcopyright}
\copyrightyear{2022}
\acmYear{2022}
\acmDOI{XXXXXXX.XXXXXXX}

\acmJournal{TOCL}
\acmVolume{37}
\acmNumber{4}
\acmArticle{111}
\acmMonth{11}

\acmPrice{15.00}
\acmISBN{978-1-4503-XXXX-X/18/06}

\begin{document}

\title{Local Search For Satisfiability Modulo Integer Arithmetic Theories}

\author{Shaowei Cai}
\authornote{The authors are listed in alphabetical order, as they all contribute significantly.}
\authornote{Corresponding author.}
\email{caisw@ios.ac.cn}
\orcid{0000-0003-1730-6922}
\author{Bohan Li}
\orcid{0000-0003-1356-6057}
\authornotemark[1]
\email{libh@ios.ac.cn}
\author{Xindi Zhang}
\orcid{0000-0001-5541-7194}
\authornotemark[1]
\email{zhangxd@ios.ac.cn}
\affiliation{%
  \institution{State Key Laboratory of Computer Science, Institute of Software, Chinese Academy of Sciences}\\
  \institution{School of Computer Science and Technology, University of Chinese Academy of Sciences}
  \city{Beijing}
  \country{China}
}

\renewcommand{\shortauthors}{Cai, Li and Zhang.}

\begin{abstract}

Satisfiability Modulo Theories (SMT) refers to the problem of deciding the satisfiability of a formula with respect to certain background first-order theories. In this paper, we focus on  Satisfiablity Modulo Integer Arithmetic, which is referred to as SMT(IA), including both linear and non-linear integer arithmetic theories. 
Dominant approaches to SMT rely on calling a CDCL-based SAT solver, either in a lazy or eager flavour. Local search,  a competitive approach to solving combinatorial problems including SAT, however, has not been well studied for SMT. 
We develop the first local-search algorithm for SMT(IA) by directly operating on variables, breaking through the traditional framework. We propose a local-search framework by considering the distinctions between Boolean and integer variables. Moreover, we design a novel operator  and scoring functions  tailored for integer arithmetic, as well as a two-level operation selection heuristic. Putting these together, we  develop a local search SMT(IA) solver called LocalSMT. Experiments are carried out to evaluate LocalSMT on benchmark sets from SMT-LIB. The results show that  LocalSMT is competitive and complementary with state-of-the-art SMT solvers, and performs particularly well on those formulae with only integer variables. 
A simple sequential portfolio with Z3 improves the state-of-the-art on satisfiable benchmark sets from SMT-LIB.

\end{abstract}

\ccsdesc[500]{Theory of computation~Logic and verification}
\ccsdesc[500]{Mathematics of computing~Randomized local search}
\ccsdesc[500]{Mathematics of computing~Solvers}


\ccsdesc[500]{Theory of computation~Linear logic}
\ccsdesc[500]{Theory of computation~Randomized local search}
\ccsdesc[300]{Theory of computation~Logic and verification}

\keywords{SMT, Local Search, Linear Integer Arithmetic, Non-linear Integer Arithmetic}

\maketitle

\section{Introduction}

Satisfiability Modulo Theories (SMT) is the  problem of deciding the satisfiability of a first-order logic formula with respect to certain background theories.
Inspired by the great success of propositional satisfiability (SAT) solving, SMT attempts to generalize the advances of satisfiability solvers from propositional logic to fragments of first-order logic.
Typical theories supported by SMT include the theories of integers, real numbers, lists, arrays and bit-vectors.
The field of SMT has seen significant progress in the past two decades. SMT solvers have become important formal verification engines, with applications in various domains.

In this paper, we focus on the theory of {\it Integer Arithmetic} (IA), consisting of arithmetic atomic formulae in the form of polynomial equalities or inequalities over integer variables.
{\it Integer Arithmetic} can be divided into two categories, namely {\it linear integer arithmetic} (LIA) and {\it non-linear integer arithmetic} (NIA), based on whether the arithmetic formulae purely consist of linear inequalities or not.
The SMT problem with the background theory of LIA and NIA is to determine the satisfiability of the Boolean combination of respective arithmetic atomic formulae and propositional variables, and referred to as SMT(LIA) and SMT(NIA).
Generally, the SMT problem of Integer Arithmetic is collectively referred to as SMT(IA).

SMT(IA) is important in software verification and  automated reasoning, since most programs use integer variables and perform arithmetic operations on them ~\cite{mccarthy1993towards}.
Specifically, SMT(LIA) has various applications in automated termination analysis ~\cite{codish2012exotic}, sequential equivalence checking ~\cite{lopes2016automatic}, and state reachability checking under weak memory models ~\cite{gavrilenko2019bmc}. 
A popular fragment of LIA, namely {\it Integer Difference Logic} (IDL), has found applications  in problems with timing-related constraints ~\cite{cotton2005satisfiability}, such as hardware models with ordered data structures ~\cite{ganai2006sdsat},  stable models computing ~\cite{janhunen2009computing}, and job shop scheduling ~\cite{roselli2018smt}.
SMT(NIA) has pervasive application in areas ranging from analysis, verification, and synthesis of software and hybrid systems~\cite{colon2003linear, platzer2009real, sankaranarayanan2004non, sankaranarayanan2004constructing} to game theory~\cite{beyene2014constraint}.

\subsection{Related Works}
Much effort has been devoted to solving SMT on integer arithmetic.
However, the approaches vary according to the corresponding theories.
Techniques for solving SMT on linear integer arithmetic (LIA and IDL) can be divided into lazy and eager approaches.
In contrast to SMT(LIA) and SMT(IDL), deciding SMT(NIA) is much more challenging, since the celebrated result of Matiyasevich ~\cite{davis1973hilbert} resolved Hilbert’s 10th problem in the negative by showing that satisfiability in the nonlinear case is undecidable.
Thus the mainstream approaches for solving SMT on non-linear integer arithmetic (NIA) are mainly incomplete.

The  most popular approach for SMT(LIA)  is the {\it lazy} approach ~\cite{sebastiani2007lazy,barrett2018satisfiability}, also known as  DPLL(T) ~\cite{nieuwenhuis2006solving}, which is a central development of SMT.
Many DPLL(T) solvers have been developed for SMT(LIA) ~\cite{dutertre2006fast,kim2007disequality,bromberger2019spass}. 
In this approach, the formula is abstracted into a  Boolean formula by replacing arithmetic atomic formulae with fresh Boolean variables. A SAT solver is used to reason about the Boolean structure and solve the  Boolean formula, while a theory solver receives assignments from the SAT solver and  solve  the conjunctions of atomic subformulae, including consistency checking of the assignments and theory-based deduction.

The effort in the lazy approach is mainly devoted to producing more effective theory solvers.
 Simplex-based linear arithmetic solvers that can be integrated efficiently in the DPLL(T) framework were studied  ~\cite{dutertre2006fast}.
A simplex-based decision procedure that minimizes the sum of infeasibilities of constraints was proposed  ~\cite{king2013simplex}.
 A theory solver made use of layering and several heuristics to achieve good performance ~\cite{griggio2012practical}.
 A theory solver called SPASS-IQ was designed to efficiently handle unbounded problems ~\cite{bromberger2016fast,DBLP:phd/hal/Bromberger19}.
According to recent SMT Competitions \footnote{https://smt-comp.github.io/}, almost all state-of-the-art SMT(LIA) (including SMT(IDL)) solvers are based on the lazy approach, including MathSAT5 ~\cite{cimatti2013mathsat5}, CVC5 ~\cite{barbosa2022cvc5}, Yices2 ~\cite{dutertre2006yices}, Z3 ~\cite{moura2008z3}, SMTInterpol ~\cite{christ2012smtinterpol} and SPASS-SATT ~\cite{bromberger2019spass}.

The other approach for linear integer arithmetic is the {\it eager} approach,  where the formula is reduced to an equi-satisfiable Boolean formula and then solved by a SAT solver.
This approach works well for SMT(IDL).
Typically, all intrinsic dependencies between integer variables are computed and encoded as Boolean constraints. 
Encoding to a Boolean formula  is done either by deriving adequate ranges for formula variables (a.k.a. small domain encoding)  ~\cite{pnueli2002small,bryant2002modeling,talupur2004range},
or by deriving all possible transitivity constraints (a.k.a per-constraint encoding)   ~\cite{strichman2002deciding}.
A hybrid method combining the strengths of two encoding schemes showed robust performance  ~\cite{seshia2003hybrid}.

The mainstream approaches for SMT(NIA) are as follows.
Most of the state-of-the-art SMT solvers supporting SMT(NIA) rely on the {\it eager} approach ({\it bit-blasting}) described in ~\cite{fuhs2007sat}, where the problem is encoded to a SAT problem, by bounding the integer variables to a limited range.
An approach relying on branch-and-bound has been explored in the context of CAD and VTS~\cite{kremer2016generalised}.
A novel method based on the MCSat (model-construction satisfiability) approach~\cite{moura2013model,jovanovic2013design} to SMT is proposed in ~\cite{jovanovic2017solving}, which reasons directly in the integers in a conflict-directed manner.
An incomplete approach is proposed to solve SMT(NIA) by reducing it to SMT(LIA) ~\cite{borralleras2012sat}.
Specifically, non-linear monomials are linearized by abstracting them with fresh variables and by performing case splitting on integer variables with finite domain.
Moreover, the approach is further improved in ~\cite{borralleras2019incomplete} by deciding which domains to enlarge in each iteration, guided by analyzing solutions to optimization problems.

Local search  is an incomplete method which plays an important role in many  combinatorial problems   ~\cite{hoos2004stochastic}.  Local search algorithms move from solution to solution in the space of candidate solutions  by applying local changes.
It has been successfully applied to the Boolean Satisfiability (SAT) problem   ~\cite{LiL12,BalintS12,cai2013local,cai2015ccanr,Biere17} and is competitive with CDCL solvers on certain types of instances.
However, very limited effort has been devoted to local search for SMT.
The idea of integrating local search solvers with theory solvers has been explored before, where a local search SAT solver WalkSAT is used to solve the Boolean skeleton of the SMT formula  ~\cite{griggio2011stochastic}.


There are two main local-search approaches for solving SMT on the theory of {\it bit vector} (BV) by operating  on the theory level.
First, a local-search algorithm was proposed in ~\cite{frohlich2015stochastic}; it performs modifications to the assignment of the inputs that can be summarized as bit-level, including bit flips, inverting assignments, incrementing and decrementing the value by 1.
It is guided by a scoring function and requires brute-force randomization and restart to avoid getting stuck in local minimums, which are successful techniques employed in local-search SAT solvers lifted to the SMT level.

In contrast, a precise propagation-based local search approach was presented in ~\cite{niemetz2016precise,niemetz2017propagation}, which is probabilistically approximately complete, indicating that it is guaranteed to find a solution theoretically.
The approach is based on propagating target values from the outputs to the input, and does not require brute-force randomization or restarts.  It lifts the concept of ATPG ~\cite{kunz1997reasoning} to the word-level.
The propagation-based local search approach is generalized in ~\cite{niemetz2020ternary} with respect to constant bits to ternary values.
The procedure is further extended to handle more bit-vector operators, and heuristics are introduced for more precise inverse value computation via bound tightening for inequality constraints.

We are not aware of any work on local search solvers for SMT on integer arithmetic theories.

\subsection{Contributions}

This work, for the first time, develops a local-search solver for SMT on the theory of quantifier-free Integer Arithmetic, denoted as SMT(IA), which directly operates on both Boolean and integer variables, breaking through the traditional approaches. The algorithm is a combination of several ideas, which  are listed as follows.

\begin{enumerate}

\item We propose a  local search framework,  which switches between two modes, namely Boolean mode and Integer mode. Each mode consists of consecutive  operations of the same type (either Boolean or integer), while keeping the assignments of variables of the other type fixed. In this way, the algorithm switches between solving subformulas consisting of only one type (either Boolean or Integer) of variables.

\item We propose a literal-level operator named {\it critical move} dedicated for Integer mode, which takes into account the literal-level information.
Specifically, given a falsified literal and an integer variable in it, a {\it critical move}  modifies the variable to the threshold value making the literal true.

\item A two-level heuristic is proposed to pick a {\it critical move}  operation. 
We introduce a special type of {\it decreasing} (meaning the operation would decrease the value of a cost function) {\it critical move} operations based on the conflict driven principle, and give a priority to such operations.

\item A fine-grained scoring function named {\it distance score} is proposed to measure the decrement on the distance of falsified clauses to satisfaction  by taking an operation. This is determined based on the cost function called the {\it distance to truth} of literals, which measures the distance of a falsified literal to becoming true.

\end{enumerate}

By putting these together, we develop a local search solver for SMT(IA) called LocalSMT.
Experiments are conducted to evaluate LocalSMT on 3 benchmark sets, including QF\_LIA, QF\_IDL and QF\_NIA benchmark sets from SMT-LIB (excluding unsatisfiable insances) \footnote{http://www.smt-lib.org/}. 
We compare our solver with state-of-the-art SMT solvers including Z3, CVC5, Yices and MathSAT5.
Moreover, we also compare LocalSMT with 5 variants of the incomplete solver dedicated for SMT(NIA) in ~\cite{borralleras2019incomplete}.
Experimental results show that LocalSMT is competitive and complementary with state-of-the-art SMT solvers. Particularly, LocalSMT is good at solving instances without Boolean variables, noting that a large portion  in SMT-LIB (78.4\% for LIA, 52.1\% for IDL and 88.7\% for NIA) belong to this type.
A simple sequential portfolio with Z3 improves the state-of-the-art on satisfiable benchmarks from SMT-LIB. 

This paper is the extended version of the conference paper of CAV'22 ~\cite{cai2022local}, where a Local Search solver aimed at SMT(LIA) was proposed.
The new contributions of this paper include extending the local search algorithm to SMT(IA), being capable of solving SMT(NIA), and conducting more experiments on SMT-LIB to evaluate the performance of LocalSMT.

\subsection{Paper Organization}
Some preliminary knowledge is introduced in Section 2.
In Section 3, we propose a local search framework for SMT(IA).
In Sections 4 and 5, we describe the two main ideas in the local search process, including the critical move operator and fine-grained scoring function, as well as a two-level heuristic for picking an operation which is a candidate variable/assignment pair.
We describe the LocalSMT algorithm in Section 6. 
Experimental results and further analysis are presented in Section 7.
Finally, we give some concluding remarks in Section 8.

\section{Preliminary}
A $monomial$ is an expression of the form $x_1^{p_1}...x_m^{p_m}$ where $m>0$, $x_i$ are variables, $p_i>0$ for all $i\in\{1...m\}$, and $x_i\not=x_j$ for all $i,j\in\{1...m\},i\not=j$. A monomial is linear if $m=1$ and $p_1=1$.

A $polynomial$ is a linear combination of monomials, that is, an arithmetic expression of the form $\sum_{i} a_i m_i$ where $a_i$ are coefficients and $m_i$ are monomials.
A polynomial with only linear monomials, which can be written as $\sum_i a_ix_i$, is $linear$, and otherwise it is $non-linear$.
A special case of non-linear polynomials is a $multilinear$ polynomial, where no variable occurs to a power of 2 or higher, meaning that each monomial is in the form of $x_1...x_m$. 

\begin{definition}
Integer Arithmetic (IA): Let $P=\{p_1,p_2,...p_n\}$ be a set of propositional (Boolean) variables and $X=\{x_1,x_2...x_m\}$ be a set of integer-valued variables.
Note that we refer to 0-arity function symbols as variables.
The integer arithmetic formulae are inductively defined.

1) $p\in P$ is a propositional atomic IA formula.

2) A polynomial inequality or equality ($\sum_{i} a_i m_i\bowtie k$) is an arithmetic atomic IA formula, where $ \bowtie\;\in \{=,\le\}$, $m_i$ are monomials consisting of integer-valued variables, $k$ and $a_i$ are constant coefficients (rationals or integers).

3) If $\psi$ and $\varphi$ are IA formulae, so are $\psi\vee \varphi$, $\psi\wedge \varphi$ and $\neg \varphi$.
\end{definition}

In the above definition, we note that with `$\leq$' and `$=$', other inequalities can also be expressed.
 Specifically, we can express $\sum_{i} a_i m_i<k$ as $\sum_{i} a_i m_i\leq k-1$, $\sum_{i} a_i m_i> k$ as $\neg (\sum_{i} (a_i m_i)\leq k)$, $\sum_{i} a_i m_i\ge k$ as $\sum_{i} (-a_i m_i)\leq (-k)$ and $(\sum_{i} a_i m_i)\not=k$ as $(\sum_{i} a_i m_i\le (k-1)\vee  \neg (\sum_{i} (a_i m_i)\leq k)$.

The arithmetic atomic formulae in {\it Linear integer arithmetic (LIA)} consist purely of linear polynomials, which can be written as $\sum_i a_i x_i\bowtie k$ ,
while those in more general {\it non-linear integer arithmetic (NIA)} are composed of polynomials.
A popular fragment of linear integer arithmetic is called {\it Integer Difference Logic} (IDL), where the arithmetic atomic formulae are in the form of $x_i-x_j\bowtie k$, where $\bowtie\in\{=,\le \}$, $x_i, x_j\in X$ and $k$ is constant.  

\begin{example}
Let $X=\{x_1,x_2,x_3,x_4,x_5\}$ and $P=\{p_1,p_2\}$ denote the sets of integer-valued and propositional variables respectively.
A typical SMT(LIA) formula $F_{LIA}$ and SMT(NIA) formula $F_{NIA}$ are shown as follows:

$F_{LIA}$: $(p_1\vee (x_1+2x_2\le2) )\wedge(p_2\vee (3x_3+4x_4+5x_5=2)\vee(-x_2-x_3\le 3)) $.

$F_{NIA}$: $(p_1\vee (x_1 x_2\le2) )\wedge(p_2\vee (3x_3^{2}x_4+4x_4+5x_5=2)\vee(-x_2-x_3\le 3)) $.

\end{example}

A literal is an atomic formula, or the negation of an atomic formula.
A $clause$ is the disjunction of a set of literals, and a formula in $conjunctive\;normal\;form$ (CNF) is the conjunction of a set of clauses.
For an SMT(IA) formula $F$, an assignment $\alpha$ is a  mapping $X\rightarrow Z$ and $P\rightarrow  \{${\it false,true}$\}$, and $\alpha(x)$ denotes the value of a variable $x$ under $\alpha$. 
A {\it complete assignment} is a mapping which assigns to each variable a value.
A literal is a true literal if it evaluates to true under the given assignment, and otherwise it is a false literal.
A clause is {\it satisfied} if it has at least one true literal, and {\it falsified} if all literals in the clause are false. A complete assignment is a {\it solution} to an SMT(IA) formula if and only if it satisfies all the clauses.

When applying local-search algorithms to solve a satisfiability problem, the search space consists of all complete assignments,  each of which is a candidate solution.
Typically, a local-search algorithm starts from a complete assignment, and iteratively modifies the assignment by changing the value of one variable to search for a satisfying assignment. 

In local search, an {\it operator} defines how to modify the candidate solution. When an operator is instantiated with a variable to yield a value to assign to this variable, we obtain an {\it operation}. For example, a standard operator for SAT is {\it flip}, which modifies the current assignment by changing the value of a Boolean variable, and  {\it flip}($x_1$) is an operation, where $x_1$ is a Boolean variable in the formula. Operators for integer variables typically have two parameters, namely the variable and the value to assign. We use $op(x,v)$ to denote the operation assigning $x$ to value $v$.

Given a formula $F$, the {\it cost} of an assignment $\alpha$, denoted as $cost(\alpha)$, is the number of falsified clauses under $\alpha$. 
In this work, We adopt the dynamic local search  (DLS) method ~\cite{dynamic,hutter2002scaling,lei2018solving}, which is a popular local search method based on the idea of modifying the evaluation function in order to prevent the search from getting stuck in local minima.
DLS typically employs weights modified during the search process.
Thus, in this work, $cost(\alpha)$ denotes the total weight of all falsified clauses under an assignment $\alpha$.
Given a formula and an assignment $\alpha$, an operation $op$ is said $decreasing$ if $cost(\alpha')<cost(\alpha)$, where $\alpha'$ is the resulting assignment by applying $op$ to $\alpha$.

\section{A Local Search Framework for SMT(IA)}

In this section, we introduce a local search framework for SMT(IA), which switches  between integer operations and Boolean operations.

\begin{figure}[ht]
\centering
\centerline{\includegraphics[scale=0.27]{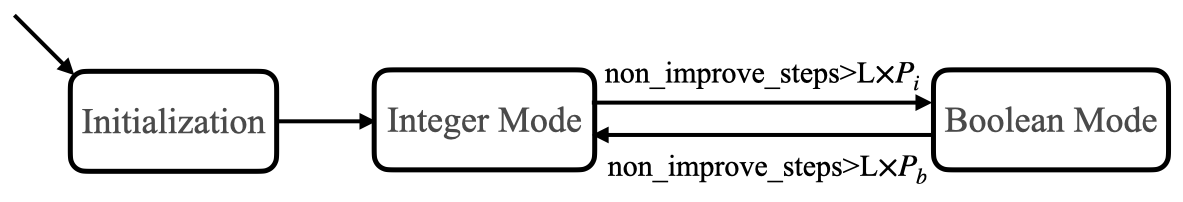}}
\caption{An SMT Local Search Framework}
\label{framework}
\end{figure}

\begin{algorithm}[!t]
\caption{Local Search of Mode $X$}
\label{ls}
\tcc{$X$ can be Integer or Boolean}
\While{{\it non\_improve\_steps} $< L\times P_X$ } {
\lIf{$\alpha$ satisfies $F$}{return $\alpha$}
\If{$\exists$ decreasing $X$ operations}{$ op:=$ a decreasing $X$ operation
}
\If{fail to find decreasing $X$ operation}{
update clause weights\;
$op:=$ an  $X$ operation from a random falsified clause containing $X$ literals\;}

perform $op$ to modify $\alpha$\;
}
\end{algorithm}

In the beginning, the algorithm generates a complete assignment $\alpha$. Then, it iteratively modifies $\alpha$ by performing operations on variables. The algorithm terminates once $\alpha$ becomes a solution to the formula, and outputs "SATISFIABLE" as well as the solution. If the algorithm fails to find a solution within the pre-set time limit, it is cut off and outputs "UNKNOWN".

As depicted in Fig. 1,  after the initialization, the algorithm works in two modes, namely Integer mode and Boolean mode. In each mode $X$ ($X$ is Integer or Boolean), an $X$ operation is picked to modify $\alpha$, where an $X$ operation refers to an operation that works on a variable of data type $X$.
The two modes switches to each other when the number of non-improving steps (denoted as $non\_improve\_steps$) of the current mode reaches a threshold. The threshold is set to $L\times P_b$ for the Boolean mode and $L\times P_i$ for the Integer mode, where $P_b$ and $P_i$ denote the proportion of Boolean and integer literals to all literals in falsified clauses, and $L$ is a parameter.
Note that $non\_improve\_steps$ is set to 0 whenever entering a mode, and then in each following step, it increases by one if $cost(\alpha')\ge cost(\alpha)$ in the current step, where $\alpha$  and $\alpha'$ denotes the assignment before and after performing the operation.

The intuitions of the two mode framework are as follows.
When all variables of one type (either Boolean or integer) are fixed, the formula is reduced to a subformula that contains only variables of the other type.
Thus, by consecutively performing $X$ ($X$ can be Boolean or Integer) operations in a certain period, the algorithm focuses on dealing with a subformula consisting of only $X$ variables.
The switching threshold is set as $L\times P_X$, as we consider that  when $X$ literals  account for larger proportion of all literals in falsified clauses, more steps should be allocated for the corresponding mode.

Consider the situation when there is no Boolean literal in falsified clauses, indicating that $P_ b=0$, but the algorithm fails to find an assignment in the following Integer mode.
Note that the set of falsified clauses is altered after the Integer mode.
If there exist Boolean literals in falsified clauses, indicating that $P_b>0$, the algorithm will switch to the Boolean mode and operates on Boolean variables.
Otherwise, the $non\_improve\_steps$ will be reset to 0, and the algorithm will enter the Integer Mode again.
Thus, The algorithm will not get stuck.

{\bf Local search in one mode:}

No matter the mode in which the algorithm works, it adopts a general procedure as described in Algorithm \ref{ls}. It prefers to pick a decreasing operation (according to some heuristic) if any. 
If the algorithm fails to find any decreasing operation, it updates clause weights by increasing the weights of falsified clauses, and then picks an $X$ operation from a random falsified clause containing $X$ literals. Note that  we can always pick a falsified clause with $X$ literals (line 7).
This is because when the algorithm works in $X$ mode, since {\it non\_improve\_steps} $< L\times P_X$, we have  $P_X>0$, and so there exists at least one falsified clause with $X$ literals.

As for clause weighting, our algorithm employs the probabilistic  version of the PAWS scheme ~\cite{cai2013local}, which is evolved from the original PAWS scheme ~\cite{thornton2004additive}. Note that the probabilistic PAWS is also adopted in previous local-search algorithm for SMT~\cite{frohlich2015stochastic}.
When the clause weighting scheme is activated, the clause weights are updated as follows.
With probability $1-sp$, the weight of each falsified clause is increased by one, and  with probability $sp$, for each satisfied clause whose weight is greater than 1, the weight is decreased by one.

\section{The Critical Move Operator and A Two-level Heuristic}

In this section, we introduce key techniques in the Integer mode of our method. We propose a novel operator called critical move.
Moreover, a two-level heuristic is proposed for choosing a critical move in the Integer mode.

A  key and basic component of a local search algorithm is the operator.
For handling Boolean variables, our algorithm adopts the typical local search  operator for SAT, namely {\it flip}, which modifies the value of a Boolean variable to the opposite of its current value (from True to False, or from False to True).
For handling integer variables, we propose an  operator called {\it critical move} which works on the literal level.
First, we introduce the {\it critical move} dedicated for SMT(LIA), and then extend it to SMT(NIA).

\subsection{Critical Move for SMT(LIA)}

Different from the Boolean operator, an integer operator has two parameters -- besides  the variable to operate on, it also needs to consider the increment (may be positive or negative) on the value.

Let us first consider a simple operator, which motivates us to propose a literal-level operator.
A simple integer operator is to modify the value of a  variable $a$ by a fixed increment $inc$, that is, $\alpha(a):=\alpha(a)\pm inc$.
The parameter $inc$ needs fine tuning.
If $inc$ is too small, it may take many iterations before making any falsified literal become true. 
If $inc$ is too big, the algorithm may even become problematic that it can never make some literals true and thus essentially unable to solve some formulae.
\begin{example}
Given a formula $F: (b-a\ge 3)\wedge (b-a\le 5)$ and the current assignment is $\alpha=\{a=0, b=0\}$.
If $inc=1$, it needs at least 3 operations to satisfy the formula.
If $inc=10$, then the formula cannot be satisfied using operations of this type, as the value of $b-a$ would be always a multiple of 10.

\end{example}
In fact, in order to avoid the case that some literals can never become true (when the $inc$ is too big), the only acceptable value of $inc$ is 1.
The main reason accounting for such a drawback is that the above operator  ignores the literal-level information.
We propose a literal-level operator for integer variables called {\it critical move}, which is defined below.

\begin{definition}
\label{cm def}
The critical move operator $cm$ has two parameters, a variable $x$ and a literal $\ell$ in the form of an arithmetic atomic formula containing $x$ or its negation, and $\ell$ is falsified under the current assignment.
$cm$ assigns  $x$ to the threshold value  making literal $\ell$ true.
\end{definition}

We first present the critical move for LIA, while the case for NIA will be introduced in the next subsection. 
Given an arithmetic atomic LIA formulae $\sum_{i}a_ix_i\bowtie k$, $\bowtie\in\{\le,=\}$, for each of the four basic forms of the falsified SMT(LIA) literal $\ell$, we denote $\Delta=\sum_{i}a_i\alpha(x_i)- k$.
The operation $cm(x,\ell)$ is described below:

\begin{itemize}
    \item $\ell: \sum_{i}a_ix_i\le k$. There exists a {\it cm} operation $cm(x_i,\ell)$ for each variable $x_i$: if the coefficient $a_i>0$, then $cm(x_i,\ell_1)$ decreases  $\alpha(x_i)$ by $\left \lceil \left| \frac{\Delta}{a_i} \right|\right \rceil$\footnote{$\left \lceil a\right \rceil$ returns the ceiling value which is the closest integer greater than or equal to a given number $a$, while  $\left| a \right|$ denotes the absolute value of $a$}; if  $a_i<0$, then  $cm(x_i,\ell_1)$ increases  $\alpha(x_i)$ by $\left \lceil \left| \frac{\Delta}{a_i} \right|\right \rceil$.
    
    \item $\ell: \neg (\sum_{i}a_ix_i\le k)$, that is, $\sum_{i}a_ix_i> k$. there exists a {\it cm} operation $cm(x_i,\ell)$ for each variable $x_i$: if the coefficient $a_i>0$, then $cm(x_i,\ell_1)$ increases  $\alpha(x_i)$ by $\left \lceil \left| \frac{1-\Delta}{a_i} \right|\right \rceil$; if  $a_i<0$, then  $cm(x_i,\ell_1)$ decreases  $\alpha(x_i)$ by $\left \lceil \left| \frac{1-\Delta}{a_i} \right|\right \rceil$.
    
    \item $\ell: \sum_{i}a_ix_i= k$.  There exists an operation $cm(x_i,\ell)$ for each variable $x_i$ with coefficient $a_i\mid\Delta$\footnote{$a\mid b$ means that $b$ is divisible by $a$.}, which increases $\alpha(x_i)$ by $-\frac{\Delta}{a_i}$. 
    
    \item $\ell: \neg (\sum_{i}a_ix_i= k)$. There exist 2 {\it cm} operations for each variable $x_i$, to increment or decrement $x_i$ by 1.
\end{itemize}

\begin{example}
Assume we are given two falsified literals $l_1: (2b-a\le -3)$ and $l_2: (5c-d+3a=5)$, and the current assignment is $\alpha=\{a=0,b=0,c=0,d=0\}$. Then $cm(a,l_1)$, $cm(b,l_1)$, $cm(c,l_2)$, and $cm(d,l_2)$ refers to assigning $a$ to 3, assigning $b$ to $-2$, assigning $c$ to 1 and assigning $d$ to $-5$ respectively. 
Note that there does not exist $cm(a,l_2)$, since $-5$ is not divisible by $3$.
\end{example}

An important property of the {\it cm} operator is that after the execution of  a {\it cm} operation, the corresponding literal must be true. 
Therefore, by picking a falsified literal  and performing a {\it cm} operation on it, we can make the literal become true.

Given the above definition of the critical move, an issue with this operator is that it may stall on equalities, when there is no such variable with coefficient $a_i\mid\Delta$ in $\ell$. 
To address this issue, in this situation, we additionally employ a simple strategy --- we pick a random variable in that literal and to increment or decrement the value of the variable by 1, which can decrease  $\left| \Delta \right|$.
For example, given an equality $2a+3b=5$, where the assignment is $\{a=0,b=0\}$, since there exists no variable with coefficient which divides $\Delta=-5$, we pick a random variable such as $a$, and increment $a$ by $1$ to decrease $\left| \Delta \right|$.

The critical move operations are analogous to update operations in other linear arithmetic model searching procedures. For example, Simplex for DPLL(T) ~\cite{dutertre2006integrating} also progresses through a sequence of candidate assignments by updating the assignment to a variable to satisfy its bound.  The significant distinction of critical moves is only updating input variables and always updating by an integral amount, as we can see from Definition 2.

\subsection{Critical Move for SMT(NIA)}

We further extend the Definition \ref{cm def} of Critical Move to SMT(NIA) literals. 
 Note that variables in NIA can have a degree greater than 1 in a polynomial.
Consider an equation with respect to the polynomial of a given literal; when all variables but one are fixed to a value, there are often multiple roots with respect to that variable (if its highest degree is greater than 1).
 This would lead to a set (instead of just one) of $cm$ operations  for such a variable w.r.t. the literal.


Given an arithmetic atomic formula $f: \sum_i a_i m_i\bowtie k$, $\bowtie\in\{\le,=\}$, where each $m_i$ is a monomial, consider the corresponding equation $\sum_i a_i m_i-k=0$. For a variable $x$ in $f$, when all variables except $x$ in the equation are substituted by its assignment, we then have an equation for $x$, written as $\sum_i a_i m_i(x)-k=0$. Let us introduce some notation before describing the critical move for NIA. 

\begin{itemize}
    \item Suppose $x$ has $n$ different roots for $\sum_i a_i m_i(x)-k=0$, we list them as $r_1<r_2<\dots<r_n$.
    \item Let $s_j\in\{+,-\}$ denote the sign of $(\sum_i a_i m_i(x)-k)$ in the interval $(r_j,r_{j+1})$, $j\in \{0,...,n\}$. 
    For the sake of convenience, we add $r_0$ and $r_{n+1}$ to denote $-\infty$ and $\infty$ respectively, although $r_0$ and $r_{n+1}$ are not roots. 
    
   
\end{itemize}


Now, we are ready to describe the critical move for NIA. 
For each of the four basic forms of a falsified SMT(NIA) literal $\ell$ and one of its variables $x$, applying the critical move operator to $x$ w.r.t. $\ell$ leads to a set of operations, depending on the roots $r_j$ and signs $s_j$ of the corresponding polynomial.
In order to distinguish from LIA, the operation set is denoted as $cm_{NIA}(x,\ell)$.
Recall that $op(x,v)$ denotes an operation assigning $x$ to value $v$.
Detailed description is as follows.

\begin{itemize}
    \item $\ell: \sum_{i}a_im_i\le k$. We have a set of critical moves as below: 

    $$cm_{NIA}(x,\ell)=  \bigcup_{j\in S^-} \{op(x,I_{min}[r_j,r_{j+1}]),op(x,I_{max}[r_j,r_{j+1}])\},$$ where $S^-=\{j\in \{0,...,n\} | s_j=$ `$-$'$\}$, $I_{min}[r_j,r_{j+1}]$ and $I_{max}[r_j,r_{j+1}]$ refers to the smallest and biggest integer in the interval $[r_j,r_{j+1}]$, respectively. Note that $r_0$ and $r_{n+1}$ refers to $-\infty$ and $+\infty$, and thus we never include $op(x,I_{min}[r_0,r1])$ or $op(x,I_{min}[r_n,r_{n+1}])$ in the set.
    
    \item $\ell: \neg (\sum_{i}a_im_i\le k)$, that is, $\sum_{i}a_ix_i> k$. We have a set of critical moves as below:
    
   $$cm_{NIA}(x,\ell)=  \bigcup_{j\in S^+} \{op(x,I_{min}(r_j,r_{j+1})),op(x,I_{max}(r_j,r_{j+1}))\},$$  
   where $S^+=\{j\in \{0,...,n\} | s_j=$ `$+$'$\}$ and $I_{min}(r_j,r_{j+1})$ and $I_{max}(r_j,r_{j+1})$ refers to the smallest and biggest integer in the interval $(r_j,r_{j+1})$, respectively.. Again, we  never include $op(x,I_{min}(r_0,r1))$ or $op(x,I_{min}(r_n,r_{n+1}))$ in the set.
    
    
    \item $\ell: \sum_{i}a_im_i= k$. 
    
    $$cm_{NIA}(x,\ell)=\{op(x,r_j)|r_j\; is\; an\; integer \}.$$
    
    
    \item $\ell: \neg (\sum_{i}a_im_i= k)$. 
    There exist 2  operations for each variable $x$, to increment or decrement $x$ by 1.
\end{itemize}

\begin{example}
Assume we are given two falsified literals $l_1: (-2bc^2+3ab+c\le -3)$ and $l_2: (5a-d^2=5)$, and the current assignment is $\alpha=\{a=1,b=1,c=1,d=1\}$.
To calculate $cm_{NIA}(c,l_1)$, the variables except $c$ in the arithmetic NIA formula are substituted by its assignment, resulting in the equation corresponding to $c$: $-2c^2+c+6=0$.
Then we need to figure out the roots of the equation, that is $r_1=-1.5,r_2=2$, and signs of corresponding intervals.
The interval corresponds to $-2c^2+c+6\le0$ is $(-\infty,-1.5]$ and $[2,\infty)$, and thus $cm_{NIA}(c,l_2)=\{op(c,2),op(c,-2)\}$.
Similarly, $cm_{NIA}(a,l_1)=\{op(a,-1)\}$, $cm_{NIA}(b,l_1)=\{op(b,-4)\}$, and $cm_{NIA}(d,l_2)=\{op(d,0)\}$.
Note that  $cm_{NIA}(a,l_2)$ contains no operation, since there is no integer root for $5a-6=0$.
\end{example}

Under circumstances where the non-linear arithmetic atomic formula is a multilinear polynomial, the critical move degenerates to the same as that in SMT(LIA).
In other words, the calculation of $cm$ operation in context of LIA is a special case of that in the context of NIA.

Note that calculating the integer solution of equations of higher degree can be time-consuming in practice, and thus in order to improve efficiency, when the variable $x$ occurs to a power of 3 or higher in the falsified literal $\ell$, the corresponding critical move $cm(x,\ell)$ will not be considered.

\subsection{A Two-level  Heuristic }

In this subsection, we propose a two-level heuristic for selecting a decreasing {\it cm} operation. We
distinguish a special type of decreasing {\it cm} operations from others, and give a priority to such operations.

From the viewpoint of algorithm design, there is a major difference between {\it cm}  and  {\it flip} operations.  A  {\it flip} operation is decreasing only if the flipping variable appears in at least one falsified clause. For a decreasing critical move operation in a falsified literal $\ell$, 
$\ell$ does not necessarily appear in any falsified clause. 
This is because $\ell$ has different variables, and changing one such variable will affect all other literals (not only $\ell$) containing the variable.

\begin{example}\label{example_two_level}  Given a formula $F=c_1 \wedge c_2 = (a-b\le0\vee b-e\le -2 \vee bd-a\le-3)\wedge(b-d\le -1)$, 
suppose the current assignment is $\alpha=\{a=1, b=1, d=1, e=1\}$, then $c_1$ is satisfied and $c_2$ is falsified.
The operation $op1=cm(b, b-e\le -2)$ and $op2=cm(b, b-d\le -1)$ refers to assigning $b$ to $-1$ and 0, respectively.
The only operation in $cm_{NIA}(b,bd-a\le-3)$, denoted as $op3$, refers to assigning $b$ to $-2$.
The literal of $op1$ and $op3$  does not appear in any falsified clause while the literal of $op2$ appears in a falsified clause $c_2$. All operations are decreasing, as either of them would make clause $c_2$ become satisfied without breaking any satisfied clause.
\end{example}

In order to find a decreasing {\it cm} operation whenever one exists, we need to scan all {\it cm} operations on false literals. Let us use $D$ to denote the candidate set of decreasing {\it cm} operations. 

\begin{itemize}
    \item For LIA: $D=\{cm(x,\ell)| \ell$ is a false literal and $x$ appears in $\ell\}$;
    \item For NIA: $D=\{op| op$ belongs to some $cm_{NIA}(x,\ell)$, where $\ell$ is a false literal and $x$ appears in $\ell\}$.
\end{itemize}

If $D=\emptyset$, there is no decreasing {\it cm} operation. 
We propose to distinguish a special subset $S\subseteq D$ from the rest of $D$, which contains {\it cm} . For LIA, $S=\{cm(x,\ell)|\ell$ appears in at least one falsified clause and $x$ appears in $\ell\}$. In the context of NIA, $S=\{op| op$ belongs to some $cm_{NIA}(x,\ell)$, where $\ell$ appears in at least one falsified clause and $x$ appears in $\ell\}$. 
Note that any {\it cm} operation in $S$ would make at least one falsified clause  become satisfied.
Based on this distinction, we propose a two-level selection heuristic as follows:

\begin{itemize}
\item The heuristic prefers to search for a decreasing {\it cm} operation from $S$.
\item If it fails to find any decreasing operation from $S$, then it searches for a decreasing {\it cm} operation from $D\setminus S$.
\end{itemize}

Besides improving the efficiency of picking a decreasing {\it cm} operation, there is an important intuition underlying this two-level heuristic. 
We prefer to pick a decreasing {\it cm} operation from $S$, because  
such operations are conflict driven, as any $cm\in S$ would force a falsified clause become satisfied. This idea can be seen as an IA version of focused local search for SAT, which has been the core idea of WalkSAT-family  SAT solvers  ~\cite{selman1993local,BalintS12,Biere17}.

\section{Scoring Functions}

Most local search algorithms employ scoring functions to guide the search ~\cite{cai2011local,cai2012configuration,cai2013local,lei2018solving}. 
We introduce two scoring functions, which are used to compare different operations and guide the local search algorithm to pick an operation to execute in each step. 

A perhaps most commonly used scoring function for SAT, denoted as {\it score}, measures the change on the cost of the assignment by flipping a variable. This scoring function indeed can be used to evaluate all types of operations as it only concerns the clauses' state (satisfied or falsified). We also employ {\it score} in our algorithm, for both {\it flip} and {\it cm} operations. Formally, the {\it score} of an operation is defined as

$$score(op)=cost(\alpha)-cost(\alpha'),$$

where $\alpha'$ is obtained from $\alpha$ by applying $op$. Note that our algorithm employs a clause weighting scheme which associates a positive integer weight to each clause, and thus the $cost$ of an assignment is the total weight of falsified clauses.
An operation $op$ is $decreasing$ if and only if $score(op)>0$.
Our algorithm prefers to choose the operation with greater $score$ in the greedy mode, for both Boolean and integer operations.

For integer operations, we propose a more fine-grained scoring function, measuring the potential benefit about pushing a falsified literal towards the direction of becoming true. Firstly, we propose a property of  literals to measure this merit.

\begin{definition}\label{def:dtt}
Given an assignment $\alpha$, for an arithmetic literal $\ell: \sum_{i}a_ix_i\leq k$, its {\it distance to truth} is $dtt(\ell,\alpha)=max\{\sum_{i}a_i\alpha(x_i)-k,0\}$.  For a Boolean literal $\ell$ and an arithmetic literal $\ell: \sum_{i}a_ix_i= k$, $dtt(\ell,\alpha)=0$  iff $\ell$ is true under $\alpha$ and  $dtt(\ell,\alpha)=1$ otherwise.
\end{definition}

Suppose the current assignment is $\alpha$, for an arithmetic literal $\ell: \sum_{i}a_ix_i\leq k$, if $ \sum_{i}a_i\alpha(x_i)>k$,  then the literal is falsified, and its $dtt$ is defined to be $\sum_{i}a_i\alpha(x_i)-k$. In this case, if we decrease the value of $x_i$ with $a_i>0$, or increase the value of $x_i$ with $a_i<0$, the $dtt$ of $\ell$ would decrease. 
When $\sum_{i}a_i\alpha(x_i)\leq k$, the literal $\ell$ is true, and thus its $dtt$ is defined to be 0.

The definition of $dtt$ for arithmetic literals resembles the violation function for constraint satisfaction problems ~\cite{hentenryck2009constraint} and the violation operator in the simplex with sum of infeasibilities for SMT ~\cite{king2013simplex}. In this work, we extend it to the clause level  to measure the distance to satisfaction of a clause in a fine-grained manner.
Based on the concept of  {\it distance to truth} of literals, we define a function to measure the distance  to satisfaction of a clause. 

\begin{definition}\label{def:dts}
Given an assignment $\alpha$, the distance to satisfaction of a clause $c$ is $dts(c,\alpha)=min_{\ell\in c}\{dtt(\ell,\alpha)\}$.
\end{definition}

According to the definition, the $dts$ is 0 for satisfied clauses, since there is at least one satisfied literal with $dtt=0$,
while $dts$ is positive for falsified clauses.
It is desirable to lead the algorithm to decrease the $dts$ of clauses. To this end, we propose a scoring function to measure the benefit of decreasing the sum of $dts$ of all clauses. Additionally, the function takes into account the clause weights as the $score$ function.

\begin{definition}\label{def:dtscore}
Given an LIA formula $F$, the {\it distance score} of an operation $op$ is defined as

$$dscore(op)=\sum_{c\in F}(dts(c,\alpha) - dts(c,\alpha'))\cdot w(c),$$
where $\alpha$ and $\alpha'$ denotes the assignment before and after performing $op$.
\end{definition}

For Boolean {\it flip} operations, $dscore$ is equal to $score$.
For integer operations, however, compared to the $score$ function which only  concerns the state (satisfied or falsified) transformations of clauses, $dscore$ is more fine-grained, as it considers the $dts$ of clauses, which are different among falsified clauses.

\begin{example}
Given a formula $F=c_1\wedge c_3\wedge c_3=(a-b\le -1)\wedge (a-c\le -5\vee a-d\le  -10)\wedge (b-c\le-5\vee b-d\le -10)$.
Suppose $w(c_1)=1, w(c_2)=2, w(c_3)=3$, and the current assignment is $\alpha=\{a=0,b=0,c=0,d=0\}$, and thus all clauses are falsified. Consider two {\it cm} operations $op1=cm(a,a-b\leq -1)$ and $op2=cm(b,a-b\leq -1)$, which assign $\alpha(a):=-1$ and $\alpha(b):=1$ respectively, leading to $\alpha'$ and $\alpha''$  respectively.
Then $score(op1)=score(op2)=1$, as they both make $c_1$ satisfied.
Also, $dts(c_2,\alpha)-dts(c_2,\alpha')=1$, and $dts(c_3,\alpha)-dts(c_3,\alpha'')=-1$, so  $dscore(op1)=(dts(c_1,\alpha)-dts(c_1,\alpha'))\cdot w(c_1)+(dts(c_2,\alpha)-dts(c_2,\alpha'))\cdot w(c_2)= 1\times 1+1\times 2=3$ and $dscore(op2)=-2$ by similar calculation.
Therefore, $op1$ is a better operation.
\end{example}

Since the computation of $dscore$ has considerable overhead, this function is only used when there is no decreasing operation.  As the reader will see in the next section, when there is no decreasing operation, only a limited number of non-decreasing operations are tried, and thus it is affordable to calculate their $dscore$.

Note that $dscore$ is not extended to SMT(NIA), since arithmetic atomic formulas in SMT(NIA) may consist of variables with exponent greater than 1, whose modification would dramatically affect the value of polynomials, and thus may mislead the search process.
Therefore, our solver does not use $dscore$ in the context of SMT(NIA).

We end this section by noting that it is usually desirable to design dedicated scoring functions for different theories. 
Besides this work, another example is the scoring function based on a recursive function for SMT(BV) ~\cite{frohlich2015stochastic}. Some local search algorithms can  even work without any scoring function, by using propagation techniques instead ~\cite{niemetz2016precise,niemetz2020ternary}.

\section{A Local Search Algorithm for SMT(IA)}

Based on the ideas in previous sections, we design a local search algorithm for SMT(IA), and implement a solver called LocalSMT, which can be used to solve SMT formulas on integer arithmetic theories, including SMT(LIA) and SMT(NIA). As described in Section 3, after the initialization, the  local search works in either Boolean or Integer mode to iteratively modify $\alpha$ until a given time limit is reached or $\alpha$ satisfies the formula $F$. This section is dedicated to the details of the initialization and the two modes of local search, as well as other optimization techniques.

{\bf Initialization}: LocalSMT generates a complete assignment $\alpha$, by assigning the variables one by one  until all variables are assigned. 
All Boolean variables are assigned to True.
As for an integer variable $x_i$, if it has upper bound $ub$ and lower bound $lb$, that is, there exist unit clauses $x_i\le ub$ and $x_i\ge lb$, it is assigned to a random value in $[lb,ub]$.
If $x_i$ only has upper (lower) bound, $x_i$ is assigned to $ub(lb)$.
Otherwise, if the variable is unbounded, it is assigned to 0.

{\bf Boolean Mode (Algorithm \ref{LS-bool})}: 
If there exist decreasing {\it flip} operations, the algorithm selects such an operation with highest $score$. 
If the algorithm fails to find any decreasing operation, it  first updates clause weights according to the weighting scheme described in Section 3. Then, it picks a random falsified clause with Boolean literals and chooses a {\it flip} operation  with greatest $score$.

{\bf Integer Mode (Algorithm \ref{LS-Int})}: 
 If there exist decreasing {\it cm} operations, the algorithm chooses a {\it cm} operation using the two-level heuristic: it first traverses falsified clauses to find a decreasing {\it cm} operation with greatest $score$ (line 3--4); if no such operation exists, it searches for a decreasing {\it cm} operation via BMS heuristic (line 5--6)  ~\cite{cai2015balance}.
Specifically, according to BMS heuristic, the algorithm samples $t$ {\it cm} operations ($t$ is a  parameter) from the false literals in satisfied clauses, and selects the decreasing one with greatest $score$.
If the algorithm fails to find any decreasing operation, it  first updates clause weights similarly to the Boolean mode. 
Then, in the context of SMT(LIA) (resp. SMT(NIA)) it picks a random falsified clause with Integer literals and chooses a {\it cm} operation  with highest $dscore$ (resp. $score$).

{\bf Restart Mechanism}: The search is restarted when the number of falsified clauses has not decreased for $MaxNoImprove$ iterations, where $MaxNoImprove$ is a parameter.
In previous local-search algorithms for SMT (in the context of BV), an exponential restart schema similar to the Luby schema for CDCL solvers~\cite{luby1993optimal} is adopted in ~\cite{frohlich2015stochastic}, while some other local search algorithms for SMT(BV) do not incorporate any restart mechanism ~\cite{niemetz2016precise,niemetz2020ternary}.

\begin{algorithm}[!t]
\caption{ Local Search of Boolean Mode}
\label{LS-bool}

\While{{\it non\_improve\_steps} $\leq L\times P_b$}{
\lIf{$\alpha$ satisfies $F$}{return $\alpha$}

{\If{$\exists$  decreasing {\it flip} operation}
{$op:=$ such an operation with the greatest $score$
}}
\Else{
update clause weights according to the PAWS scheme\;
$c:=$ a random falsified clause with Boolean variables\;
$op:=$ a {\it flip} operation in $c$ with the greatest $score$\;
}
$\alpha:=\alpha$ with $op$ performed\; 
}

\end{algorithm}

\begin{algorithm}[!t]
\caption{ Local Search of Integer Mode}
\label{LS-Int}
\While{{\it non\_improve\_steps} $\leq L\times P_i$}{
\lIf{$\alpha$ satisfies $F$}{return $\alpha$}
{\If{$\exists$  decreasing {\it cm} operation in falsified clauses}{
$op:=$ such an operation with the greatest $score$}
\ElseIf{$\exists$ decreasing {\it cm} operation in satisfied clauses}
{$op:=$ such an operation with greatest $score$}
}
\Else{
update clause weights according to the PAWS scheme\;
$c:=$ a random falsified clause with integer variables\;
$op:=$ a {\it cm} operation in $c$ with the greatest $dscore(score)$\;
}
$\alpha:=\alpha$ with $op$ performed\; 
}

\end{algorithm}

{\bf Forbidding Strategies:}
Local search methods tend to be stuck in suboptimal regions.
To address the cycle phenomenon (i.e revisiting some search regions), we employ a popular forbidding strategy, called the tabu strategy  ~\cite{glover1998tabu}.
After an operation is executed, the tabu strategy  forbids the reverse operations in the following $tt$ iterations, where  $tt$ is a parameter usually called {\it tabu tenure}.
The tabu strategy can be directly applied in LocalSMT. (1) If a {\it flip} operation is performed to flip a  Boolean variable, then the variable  is forbidden to flip in the following $tt$ iterations. (2) If a {\it cm} operation that increases (decreases, resp.) the value of an integer variable $x$ is performed,  then it is forbidden to decrease (increase, resp.) the value of $x$ in the following $tt$ iterations.

This might be the first time a  forbidding strategy is applied in local-search algorithms for SMT. 

{\bf Discussion of Completeness:}
Our algorithm is not complete in the sense of probabilistically approximately complete (PAC) based on the following reasons.
First, the $cm$ operation always assigns the variable to the threshold value to satisfy a literal, and thus it can miss the possible satisfying solution where variables are not necessarily on the threshold to satisfy a literal.
Moreover, in the case of NIA, when the variable occurs to a power of 3 or higher in a falsified literal, the corresponding $cm$ operation is not considered, and thus the search space cannot be thoroughly traversed.


\section{Experiments}

We carry out experiments to evaluate our solver LocalSMT on 3 benchmark sets from SMT-LIB, and compare it with state-of-the-art SMT solvers.
We also compare our solver with incomplete solvers dedicated for SMT(NIA).
Moreover, we combine LocalSMT  with Z3 to obtain a sequential portfolio solver, which shows further improvement.
Additionally, experiments are conducted to analyze the effectiveness of the proposed ideas.

\subsection{Experiment Preliminaries}

{\bf Implementation: }
LocalSMT is implemented in C++ and compiled by g++ with '-O3' option.
There are 5 parameters in LocalSMT:
 $L$ for switching phases, $tt$ for the tabu scheme, $MaxNoImprove$ for restart, $t$ (the number of samples) for the BMS heuristic and $sp$ (the smoothing probability) for the PAWS scheme.
The parameters are tuned according to suggestions from the literature and our preliminary experiments on 20\% sampled instances, and are set as follows:
$L=20$, $t=45$, $tt=3+rand(10)$, $MaxNoImprove=500000$ and $sp=0.0003$ for all benchmarks. 

{\bf Competitors: }
We compare LocalSMT with 4 state-of-the-art SMT solvers according to SMT-COMP 2021\footnote{https://smt-comp.github.io/2021}, namely
MathSAT5 (version 5.6.8),
CVC5 (version 1.0.2),
Yices2 (version 2.6.4),
and Z3 (version 4.8.17), which are the union of the top 4 solvers (excluding portfolio solvers) of QF\_NIA, QF\_LIA and QF\_IDL tracks.
In the context of SMT(NIA), we also compare LocalSMT with 5 variants of the incomplete solvers dedicated to SMT(NIA) in ~\cite{borralleras2019incomplete}, namely bcl-maxsmt, bcl-ninc-cores, bcl-ninc, bcl-omt and bcl-cores.
The binaries of all competitors are downloaded from their websites.


{\bf Benchmarks: } Our experiments are carried out on 3 benchmark sets from SMT-LIB.
As LocalSMT is an incomplete solver, UNSAT instances are excluded.

\begin{itemize}
    \item SMTLIB-LIA: This benchmark set consists of  8512 unknown and satisfiable SMT(LIA)  instances from SMT-LIB\footnote{https://clc-gitlab.cs.uiowa.edu:2443/SMT-LIB-benchmarks/QF\_LIA}.
    \item SMTLIB-IDL: This benchmark set consists of 1611 unknown and satisfiable SMT(IDL)  instances from  SMT-LIB\footnote{https://clc-gitlab.cs.uiowa.edu:2443/SMT-LIB-benchmarks/QF\_IDL}.
    \item SMTLIB-NIA: This benchmark set consists of 18419  unknown and satisfiable SMT(NIA) instances from SMT-LIB\footnote{https://clc-gitlab.cs.uiowa.edu:2443/SMT-LIB-benchmarks/QF\_NIA}.
\end{itemize}

Instances from SMTLIB-LIA, SMTLIB-IDL and SMTLIB-NIA benchmark sets are divided into two categories depending on whether they contain Boolean variables. 
From the viewpoint of algorithm design, there is a major difference between the operations on Boolean and integer variables.
We observe that instances containing only integer variables make up a large proportion, amount to 78.4\%, 52.1\% and 88.7\%, in these three benchmark sets.

Since names of recently uploaded benchmarks are too long, in order to save space, they are renamed by only using the type and uploaded year.

{\bf Experiment Setup: }
All experiments are carried out on a server with Intel Xeon Platinum 8153 2.00GHz and 2048G RAM under the system CentOS 7.9.2009.
Each solver is executed with a cutoff time of 1200 seconds (as in the SMT-COMP) for each instance in SMT-LIB, as they contain sufficient instances.
``\#inst'' denotes the number of instances in each family.
We compare the number of instances where an algorithm finds a model (``\#solved''), as well as the run time.
The bold value in table emphasizes the solver with greatest ``\#solved''.

Our solver is uploaded as a part of a derived solver called Z3++\footnote{https://github.com/shaowei-cai-group/z3pp}, which won the Biggest Lead and Largest Contribution Gold Medal in Model-validation Track of SMT-COMP 2022.

\begin{table}[]
\centering
\caption{Results on instances from SMTLIB-LIA.
}
\label{tbl:lia}
\setlength{\tabcolsep}{2.1mm}\scalebox{1.0}{
\begin{tabular}{@{}llllllll@{}}
\toprule
Family                                                                      & Type                  & \#inst & MathSAT5     & CVC5         & Yices2       & Z3           & LocalSMT        \\ \midrule
\multirow{15}{*}{\begin{tabular}[c]{@{}l@{}}Without\\ Boolean\end{tabular}} & Bromberger(2018)   & 631    & 538 & 425          & 358          & 532          & \textbf{581}           \\
                                                                            & bofill-scheduling     & 407    & \textbf{407} & 402          & \textbf{407} & 405          & 391           \\
                                                                            & CAV\_2009\_benchmarks & 506    & \textbf{506} & 498          & 396          & \textbf{506} & \textbf{506}  \\
                                                                            & check                 & 1      & 1            & 1            & 1            & 1            & 1             \\
                                                                            & convert               & 280    & 273          & 205          & 186          & 184          & \textbf{279}  \\
                                                                            & dillig                & 230    & \textbf{230} & \textbf{230} & 200          & \textbf{230} & \textbf{230}  \\
                                                                            & miplib2003            & 16     & 10           & 9            & 11           & 8            & \textbf{13}   \\
                                                                            & pb2010                & 41     & 14           & 5            & 21           & \textbf{33}  & 28            \\
                                                                            & prime-cone            & 19     & 19           & 19           & 19           & 19           & 19            \\
                                                                            & RWS                   & 20     & 11           & 13           & 11           & \textbf{14}  & 12            \\
                                                                            & slacks                & 231    & 230          & \textbf{231} & 161          & 230          & \textbf{231}  \\
                                                                            & SMPT(2022)         & 4285 & 4200  & 4201         & \textbf{4220} & \textbf{4220}        & 4184 \\
                                                                            & wisa                  & 3      & 3            & 3            & 3            & 3            & 3             \\
                                                                            &                       &        &              &              &              &              &              \\
                                                                            & Total                 & 6670   & 6442         & 6242         & 5994         & 6385         & \textbf{6478} \\ \midrule
\multirow{14}{*}{\begin{tabular}[c]{@{}l@{}}With \\ Boolean\end{tabular}}   & arctic-matrix         & 100    & 43           & 26           & 59           & 47           & \textbf{77}   \\
                                                                            & Averest               & 9      & \textbf{9}   & \textbf{9}   & \textbf{9}   & \textbf{9}   & 7             \\
                                                                            & calypto               & \textbf{24}      & \textbf{24}          & \textbf{24}           & \textbf{24}           & \textbf{24}           & 21             \\
                                                                            & CIRC                  & 18     & \textbf{18}  & \textbf{18}  & \textbf{18}  & \textbf{18}  & 3             \\
                                                                            & cmodelsdiff(2019)      & 144    & 94           & \textbf{95}  & \textbf{95}  & \textbf{95}  & 51            \\
                                                                            & ezsmt(2019)            & 108    & \textbf{84}  & 79           & 81           & 81           & 54            \\
                                                                            & Dartagnan(2021)    & 47     & 22           & 22           & \textbf{23}  & \textbf{23}  & 2             \\
                                                                            & fft    & 5      & 3            & 3            & 3            & 3            & 3            \\
                                                                            & mathsat               & 21     & \textbf{21}  & \textbf{21}  & \textbf{21}  & \textbf{21}  & 13            \\
                                                                            & nec-smt               & 1256     & 1244         &  425         & \textbf{1256}& 1242         & 581            \\
                                                                            & RTCL                  & 2      & 2            & 2            & 2            & 2            & 2             \\
                                                                            & tropical-matrix       & 108    & 55           & 42           & 71           & 52           & \textbf{98}   \\
                                                                            &                       &        &              &              &              &              &              \\
                                                                            & Total                 & 1842    & 1619         & 766          & \textbf{1662} & 1617          & 912           \\ \bottomrule
\end{tabular}
}
\end{table}

\begin{figure}
     \centering
     \begin{subfigure}[b]{0.45\textwidth}
         \centering
         \includegraphics[width=5cm]{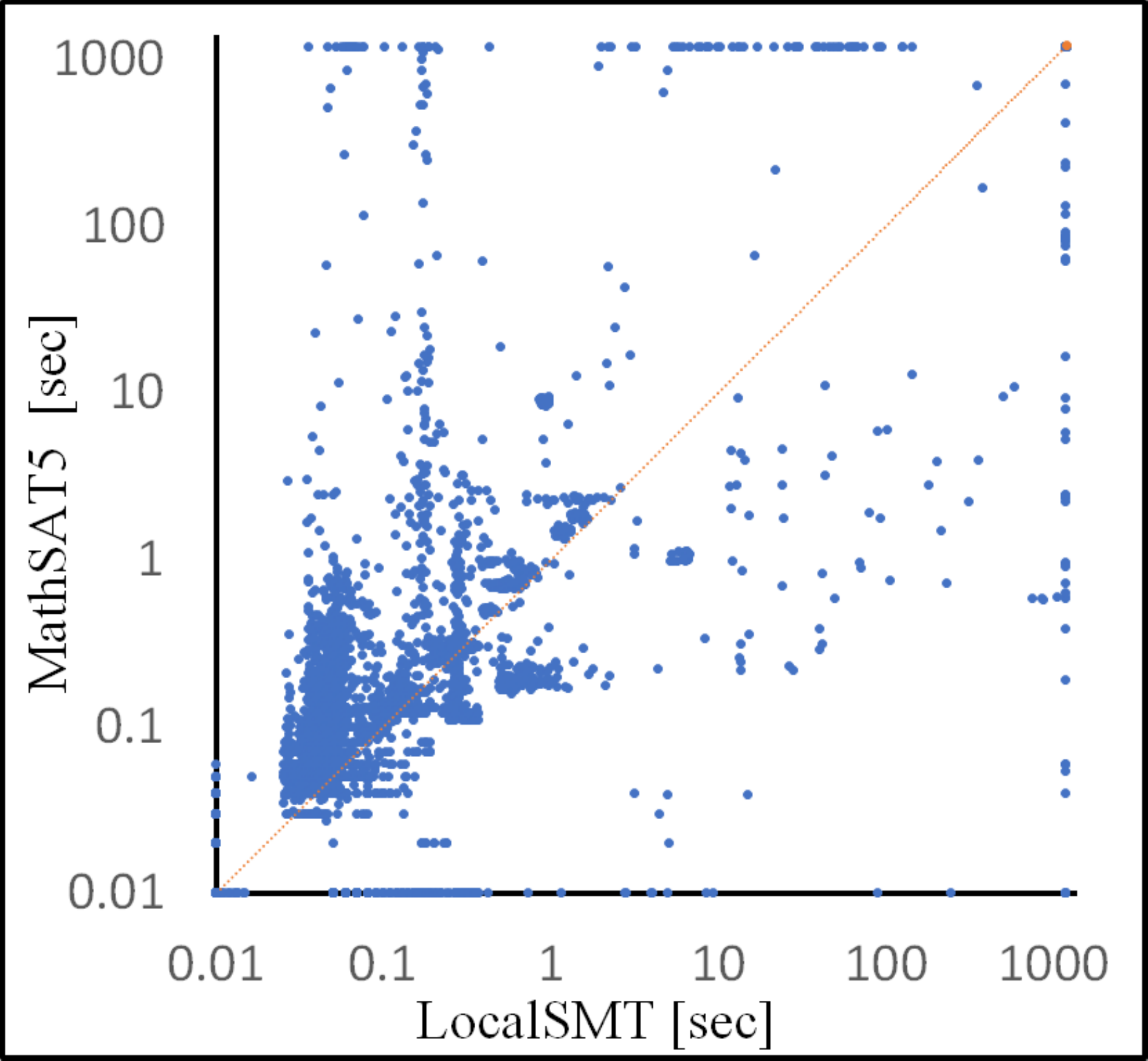}
         \caption{ Comparing with MathSAT5}
     \end{subfigure}
     \begin{subfigure}[b]{0.45\textwidth}
         \centering
         \includegraphics[width=5cm]{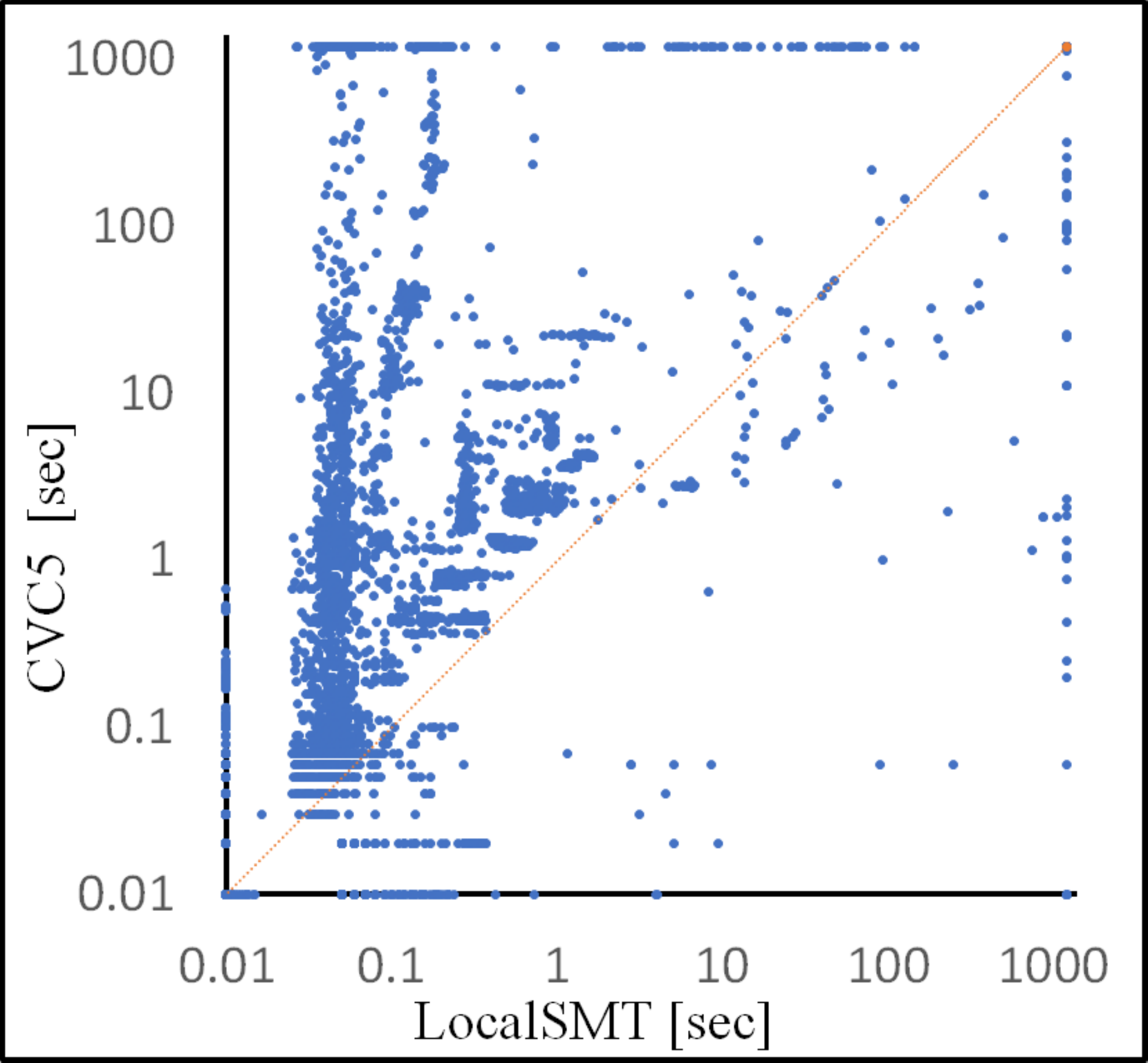}
         \caption{ Comparing with CVC5}
     \end{subfigure}

     \begin{subfigure}[b]{0.45\textwidth}
         \centering
         \includegraphics[width=5cm]{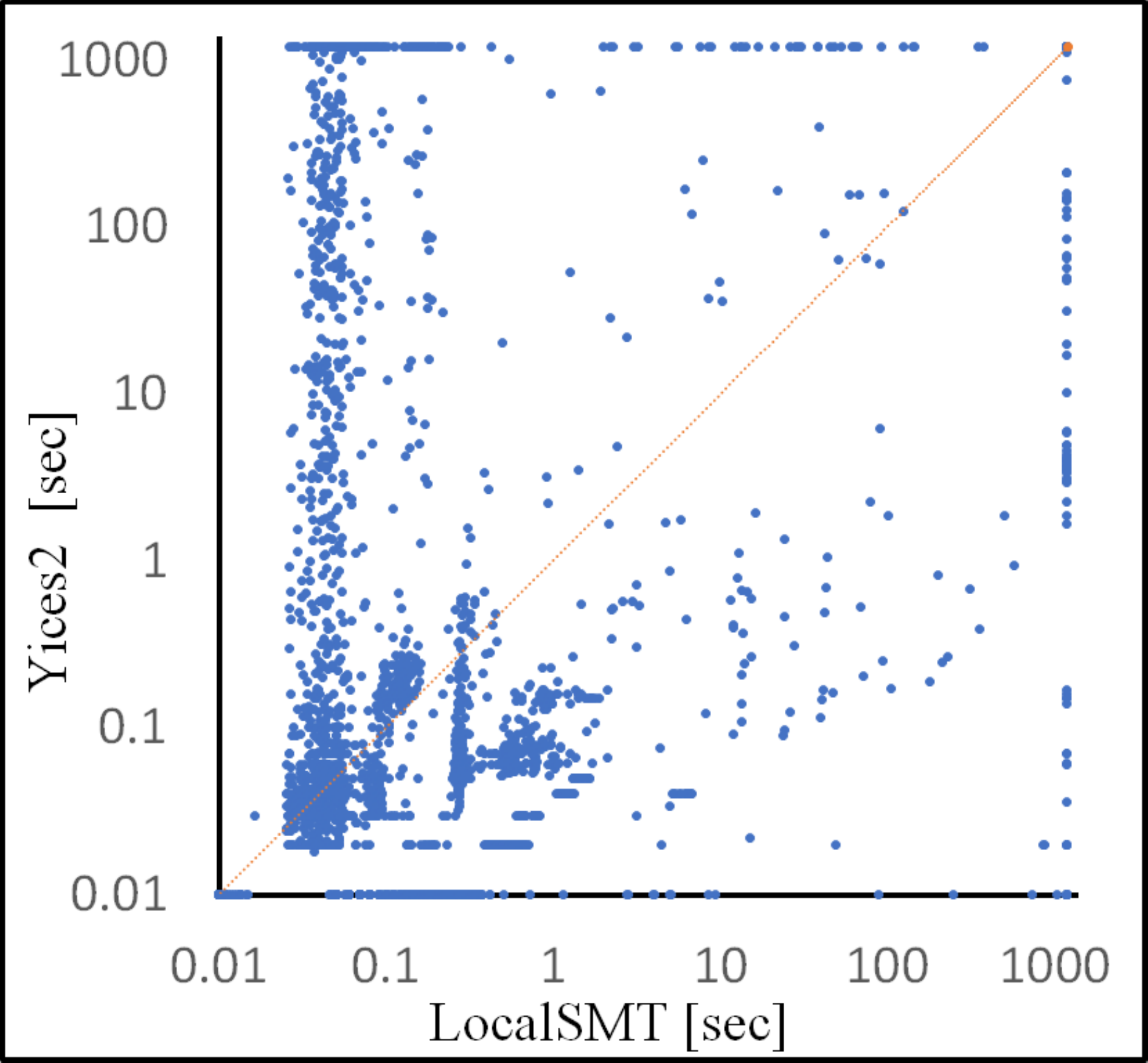}
         \caption{ Comparing with Yices2}
     \end{subfigure}
     \begin{subfigure}[b]{0.45\textwidth}
         \centering
         \includegraphics[width=5cm]{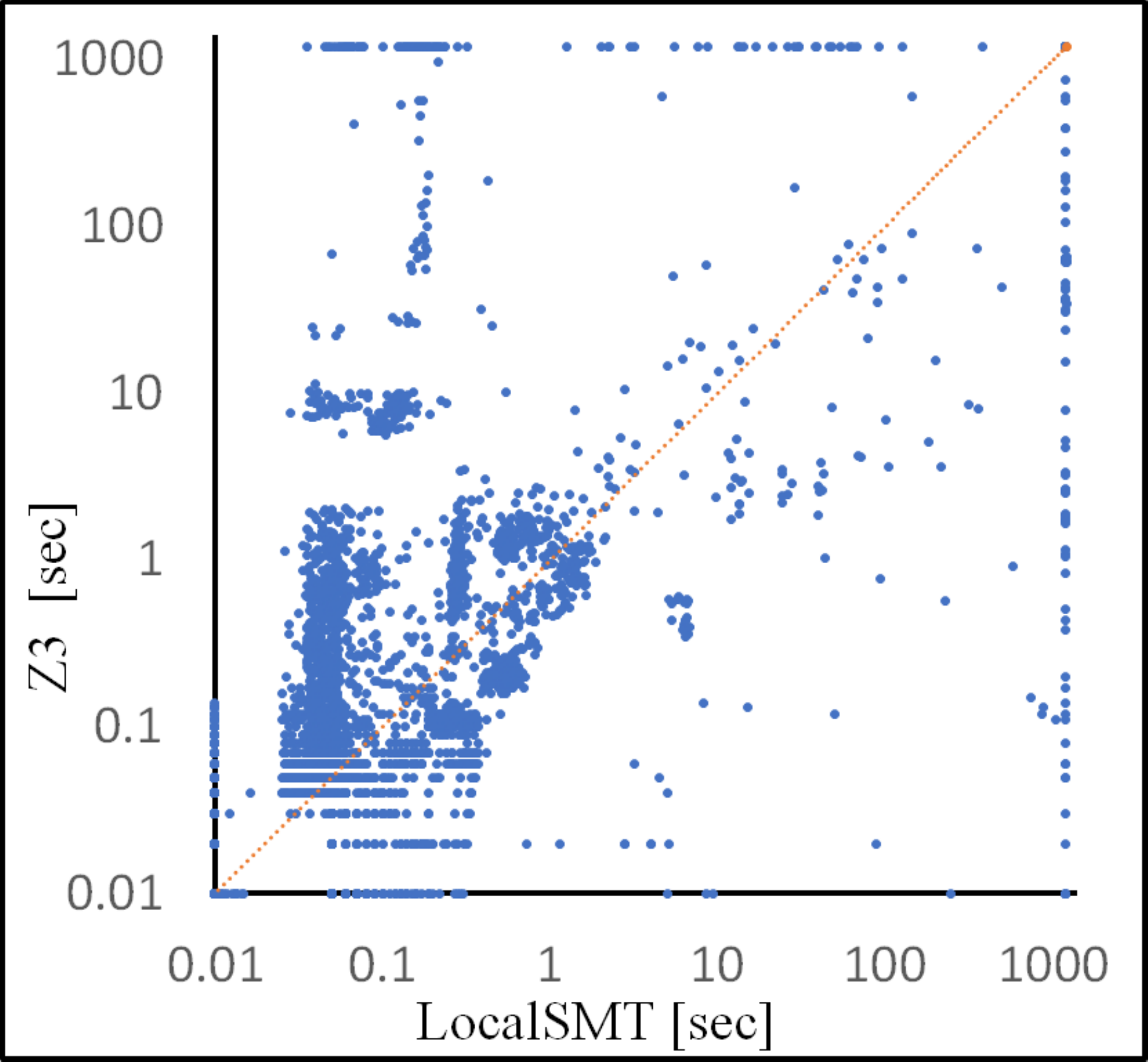}
         \caption{ Comparing with Z3}
     \end{subfigure}
        \caption{Run time comparison on Without Boolean category of SMTLIB-LIA}
        \label{fig:lia}
\end{figure}

\subsection{ Results on SMTLIB-LIA (Table \ref{tbl:lia}, Fig. \ref{fig:lia})} We organize the results into two categories: instances Without Boolean variables, and instances With Boolean variables.
LocalSMT outperforms its competitors on the Without Boolean category, solving  6478 out of the 6670 instances. We also present the run time comparisons between LocalSMT and each competitor on the Without Boolean category of SMTLIB-LIA benchmark set in Figure \ref{fig:lia}.

As for the With Boolean category, the performance of LocalSMT is overall worse than its competitors, but still comparable.
A possible explanation is that as local search SAT solvers, LocalSMT is not good at exploiting the relations among Boolean variables.
Nevertheless, LocalSMT has obvious advantage in ``tropical-matrix'' and ``arctic-matrix'' instances, which are industrial instances from automated program termination analysis ~\cite{codish2012exotic}.
LocalSMT can exclusively solve 18 and 27 instances on these two types, showing its complementary performance compared to CDCL(T) solvers.

\begin{table}[]
\centering
\caption{ Results on instance from SMTLIB-IDL.}
\label{tbl:idl}
\setlength{\tabcolsep}{2.6mm}\scalebox{1}{
\begin{tabular}{@{}llllllll@{}}
\toprule
Family                                                                    & Type            & \#inst & MathSAT      & CVC5         & Yices2       & Z3           & LocalSMT      \\ \midrule
\multirow{11}{*}{\begin{tabular}[c]{@{}l@{}}Without\\ Boolean\end{tabular}} 
                                                                        & Bouvier(2021)           & 100    & 4           & 42          & 38           & \textbf{56}  & 41           \\
                                                                        & DTP                        & 32     & 32          & 32          & 32           & 32           & 32           \\
                                                                        & job\_shop                  & 108    & 47          & 63          & 74           & 74           & \textbf{76}  \\
                                                                        & jobshop(2022)   & 119    & 26          & 28          & 48           & 44           & \textbf{75}  \\
                                                                        & n\_queen                   & 97     & 64          & \textbf{97} & \textbf{97}  & 92           & \textbf{97}  \\
                                                                        & RVpredict(2022) & 15     & \textbf{15} & \textbf{15} & \textbf{15}  & 5            & \textbf{15}  \\
                                                                        & schedulingIDL              & 247    & 105         & 162         & \textbf{247} & \textbf{247} & \textbf{247} \\
                                                                        & super\_queen               & 91     & 59          & 90          & \textbf{91}  & \textbf{91}  & \textbf{91}  \\
                                                                        & toroidal\_bench            & 32     & 11          & 10          & 12           & 12           & \textbf{13}  \\
                                                                        &                           &        &             &             &              &              &           \\
                                                                        & total                      & 841    & 363         & 539         & 654          & 653          & \textbf{687} \\ \midrule
\multirow{13}{*}{\begin{tabular}[c]{@{}l@{}}With\\ Boolean\end{tabular}}  & asp             & 379    & 147          & 212          & 284          & \textbf{291} & 27           \\
                                                                          & Averest         & 157    & \textbf{157} & \textbf{157} & \textbf{157} & \textbf{157} & 120          \\
                                                                          & bcnscheduling   & 6      & 3            & \textbf{4}   & \textbf{4}   & \textbf{4}   & \textbf{4}   \\
                                                                          & fuzzy-matrix    & 15     & 0            & 0            & 0            & 0            & \textbf{1}   \\
                                                                          & mathsat         & 16     & \textbf{16}  & \textbf{16}  & \textbf{16}  & \textbf{16}  & 11           \\
                                                                          & parity          & 136    & 130          & \textbf{136} & \textbf{136} & \textbf{136} & \textbf{136} \\
                                                                          & planning        & 2      & \textbf{2}   & \textbf{2}   & \textbf{2}   & \textbf{2}   & 0            \\
                                                                          & qlock           & 36     & \textbf{36}  & \textbf{36}  & \textbf{36}  & \textbf{36}  & 0            \\
                                                                          & RTCL            & 4      & 4            & 4            & 4            & 4            & 4            \\
                                                                          & sal             & 10     & \textbf{10}  & \textbf{10}  & \textbf{10}  & \textbf{10}  & 8            \\
                                                                          & sep             & 9      & \textbf{9}   & \textbf{9}   & \textbf{9}   & \textbf{9}   & 8            \\
                                                                          &                 &        &              &              &              &              &              \\
                                                                          & Total           & 770    & 514          & 586 & 658          & \textbf{665}          & 319          \\ \bottomrule
\end{tabular}
}
\end{table}

\begin{figure}[t]
\centering
\centerline{\includegraphics[scale=0.3]{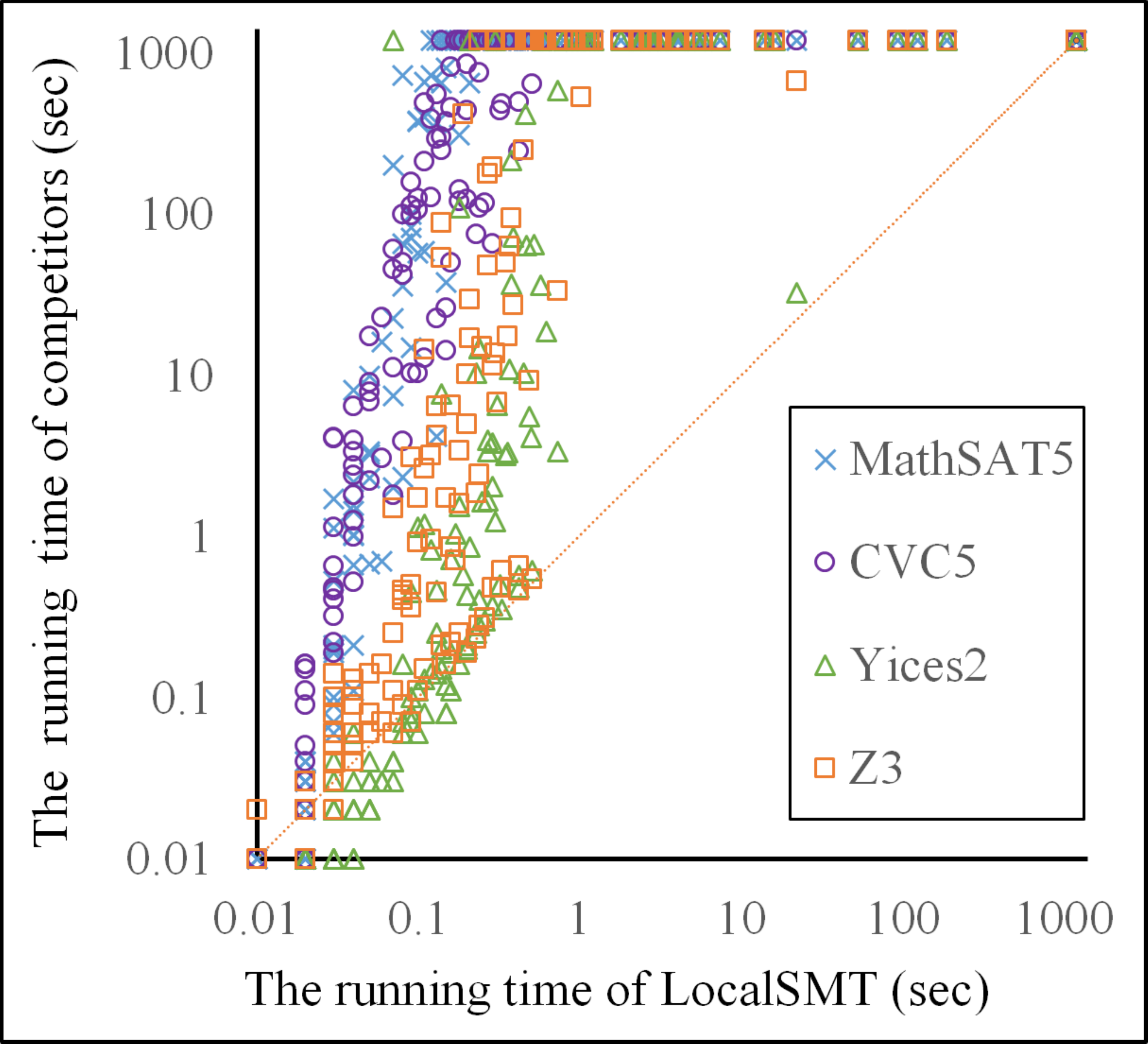}}
\caption{Run time comparison on job shop scheduling instances.}
\label{new jsp}
\end{figure}

\begin{figure}
     \centering
     \begin{subfigure}[b]{0.45\textwidth}
         \centering
         \includegraphics[width=5cm]{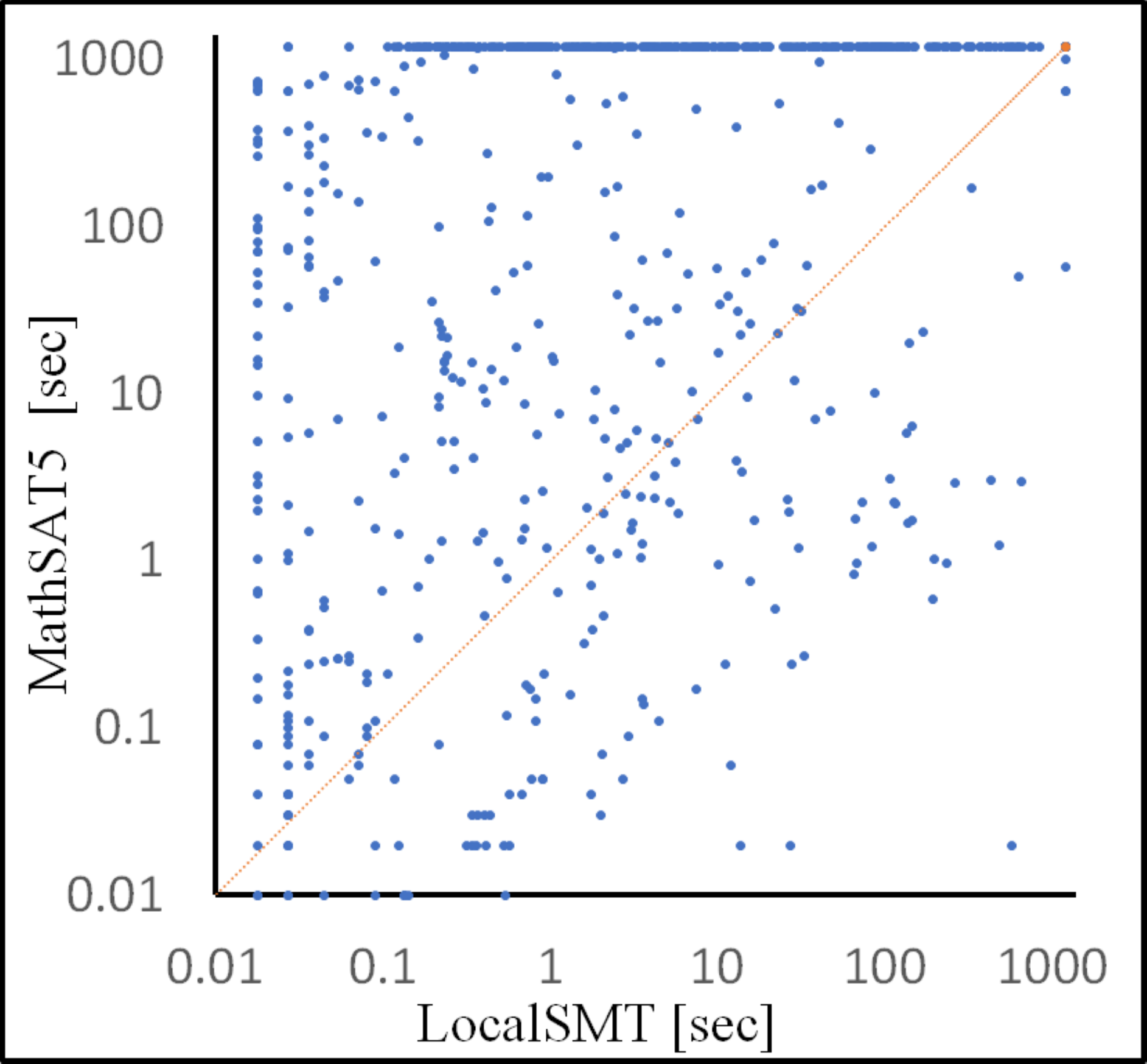}
         \caption{ Comparing with MathSAT5}
     \end{subfigure}
     \begin{subfigure}[b]{0.45\textwidth}
         \centering
         \includegraphics[width=5cm]{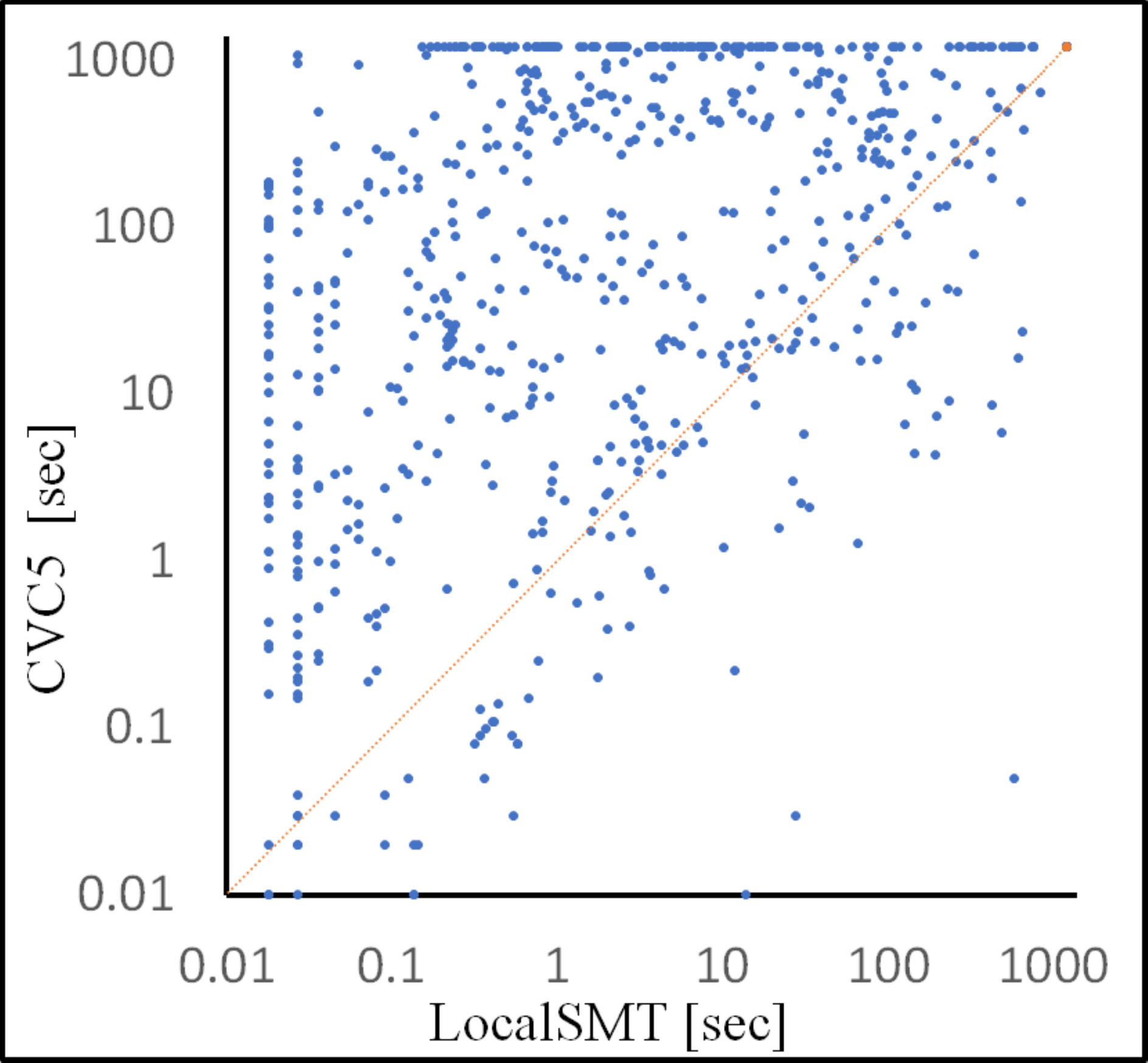}
         \caption{ Comparing with CVC5}
     \end{subfigure}

     \begin{subfigure}[b]{0.45\textwidth}
         \centering
         \includegraphics[width=5cm]{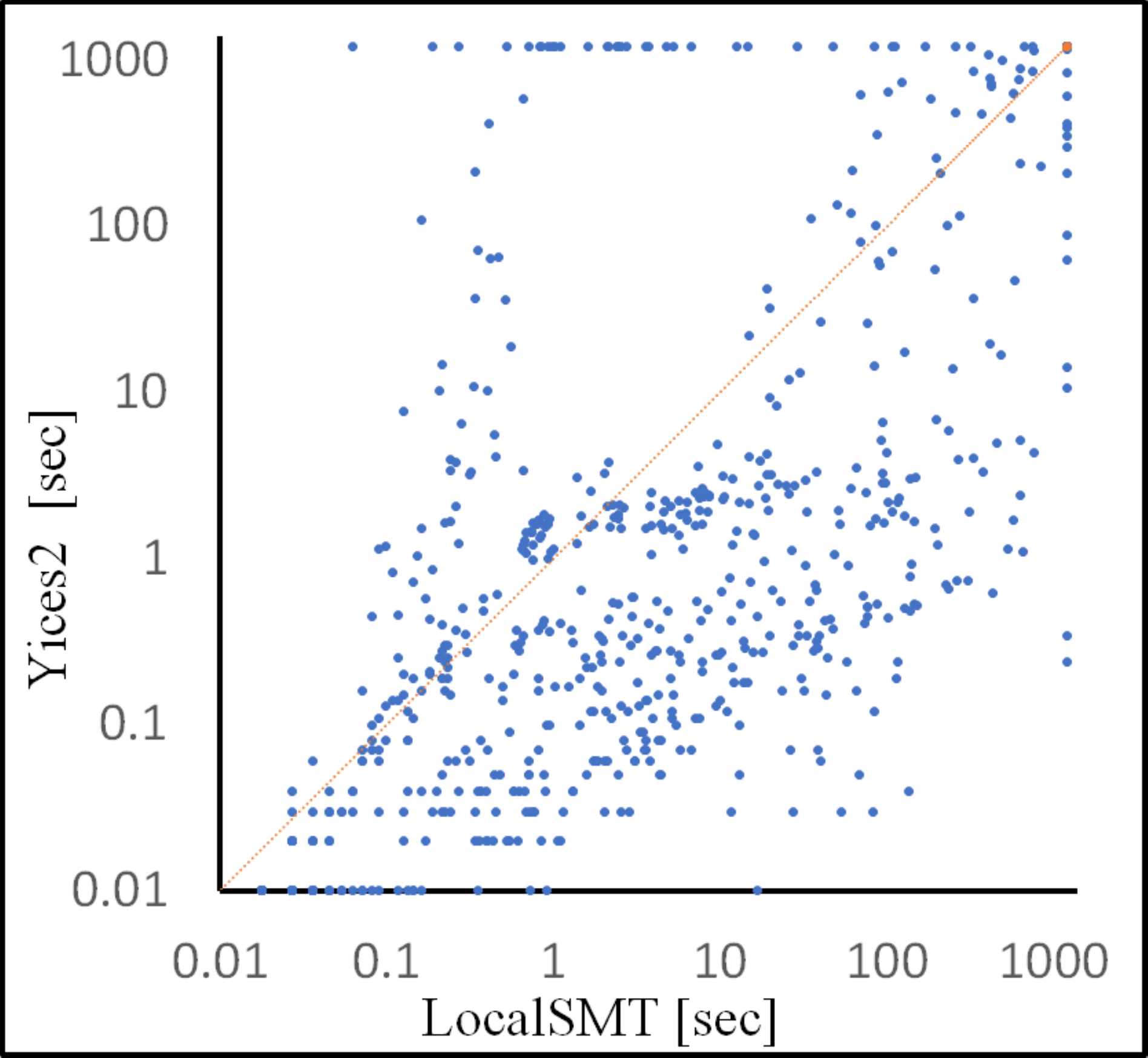}
         \caption{ Comparing with Yices2}
     \end{subfigure}
     \begin{subfigure}[b]{0.45\textwidth}
         \centering
         \includegraphics[width=5cm]{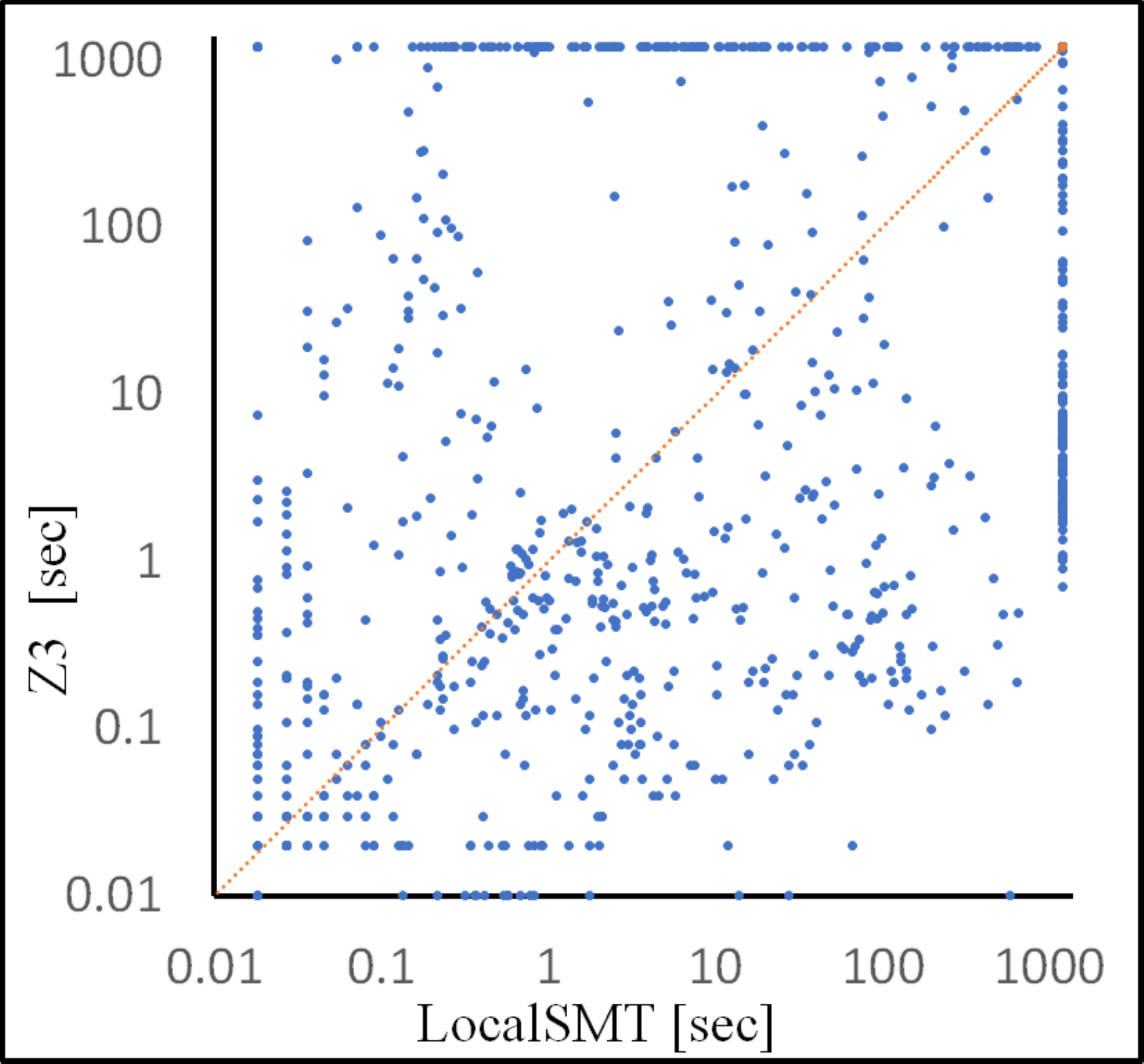}
         \caption{ Comparing with Z3}
     \end{subfigure}
        \caption{Run time comparison on Without Boolean category of SMTLIB-IDL}
        \label{fig:idl}
\end{figure}

\subsection{\bf Results on SMTLIB-IDL Benchmark set (Table \ref{tbl:idl}, Fig. \ref{fig:idl})}
Similar to the case for SMTLIB-LIA, our local search solver shows the best performance on SMT(IDL) instances without Boolean variables (solving 687 out of the 841 instances), which can be seen from Table \ref{tbl:idl} and Fig. \ref{fig:idl}.
LocalSMT works particularly well on instances from ``job\_shop'' and ``jobshop(2022)'', which are job shop scheduling problem adopting the encoding method in ~\cite{kim2007disequality}.
The run time comparison on these types of instances is presented in Fig. \ref{new jsp}.

However, LocalSMT performs worse than its competitors on those With Boolean variables.
Overall, LocalSMT cannot rival its competitors on this benchmark set, but works particularly well on the instances without Boolean variables.


\begin{table}[]
\centering
\caption{Results on instances from SMTLIB-NIA.
}
\label{tbl:nia}
\setlength{\tabcolsep}{2.1mm}\scalebox{1.0}{
\begin{tabular}{@{}llllllll@{}}
\toprule
Family                           & Type                & \#inst & MathSAT5    & CVC5        & Yices2     & Z3            & LocalSMT          \\ \midrule
\multirow{8}{*}{Without Boolean} & AProVE              & 1676   & 1647        & 1373        & 1593       & \textbf{1658} & 1627           \\
                                 & CInteger            & 1134   & 717         & 323         & 520        & 771           & \textbf{801}   \\
                                 & ITS                 & 12414  & 7785        & 5502        & 6816       & 8540          & \textbf{9449}  \\
                                 & LassoRanker         & 4      & 4           & 4           & 4          & 4             & 4              \\
                                 & leipzig             & 162    & 128         & 90          & 101        & \textbf{159}  & 154            \\
                                 & mcm                 & 181    & 13          & 16          & 11         & 15            & \textbf{97}    \\
                                 & MathProblems(2022)        & 868    & 203         & 227         & 112        & 659           & 0             \\
                                 &                     &        &             &             &            &               &       \\
                                 & Total               & 16439  & 10497       & 7535        & 9157       & 11806         & \textbf{12132} \\ \midrule
\multirow{7}{*}{With Boolean}    & calypto             & 80     & 78          & 79          & 79         & \textbf{80}   & 32             \\
                                 & Dartagnan(2021)  & 33     & \textbf{13} & \textbf{13} & 9          & \textbf{13}   & 0              \\
                                 & ezsmt(2019)          & 8      & \textbf{8}  & \textbf{8}  & \textbf{8} & \textbf{8}    & 0              \\
                                 & SAT14               & 1853   & 1801        & 1802        & 1840       & \textbf{1852} & 1633           \\
                                 & UltimateLassoRanker & 6      & \textbf{6}  & \textbf{6}  & \textbf{6} & \textbf{6}    & 4              \\
                                 &                     &        & \textbf{}   & \textbf{}   & \textbf{}  & \textbf{}     &                \\
                                 & Total               & 1980   & 1906        & 1908        & 1942       & \textbf{1959} & 1669           \\ \bottomrule
\end{tabular}
}
\end{table}

\begin{figure}
     \centering
     \begin{subfigure}[b]{0.45\textwidth}
         \centering
         \includegraphics[width=5cm]{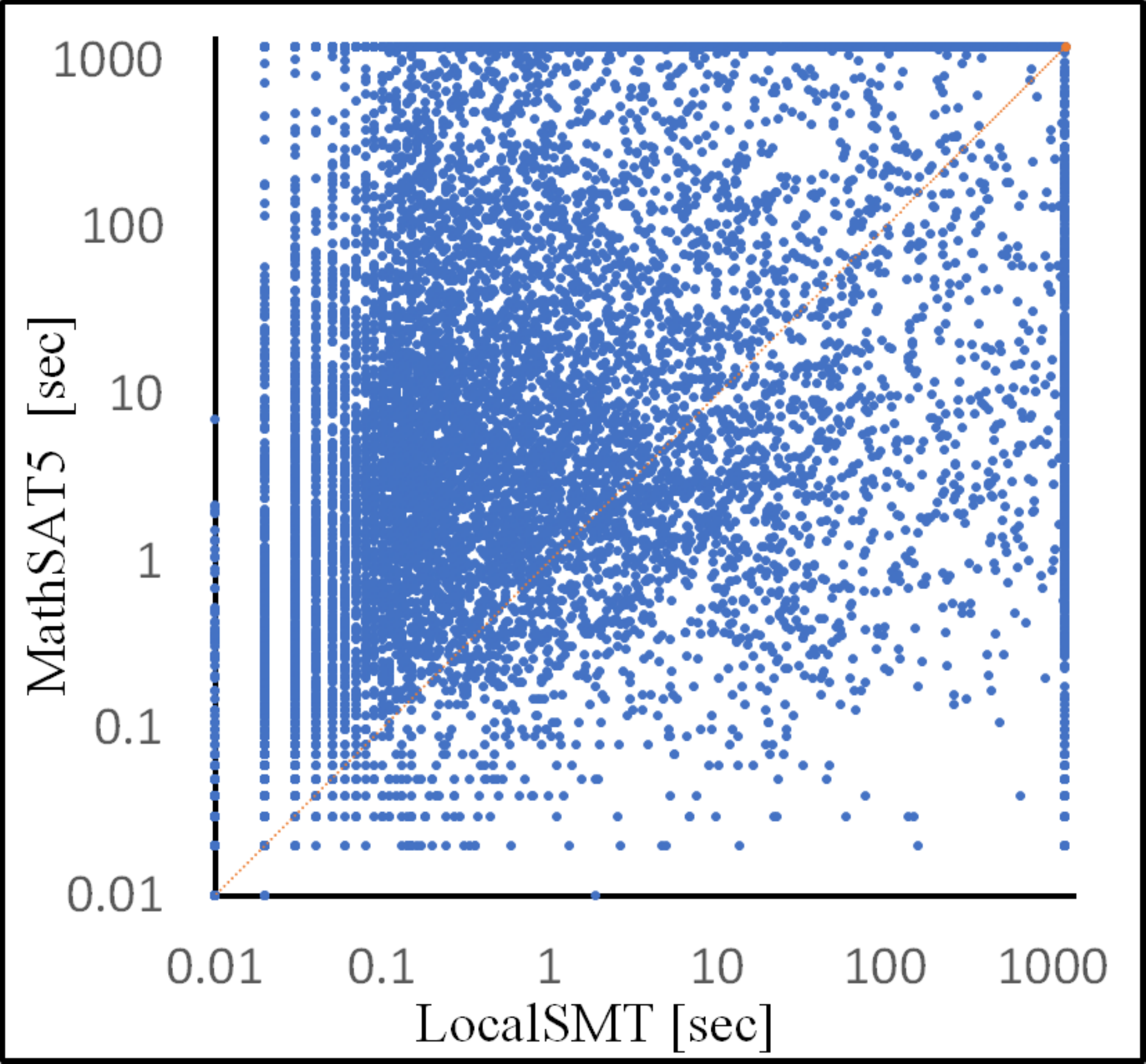}
         \caption{ Comparing with MathSAT5}
     \end{subfigure}
     \begin{subfigure}[b]{0.45\textwidth}
         \centering
         \includegraphics[width=5cm]{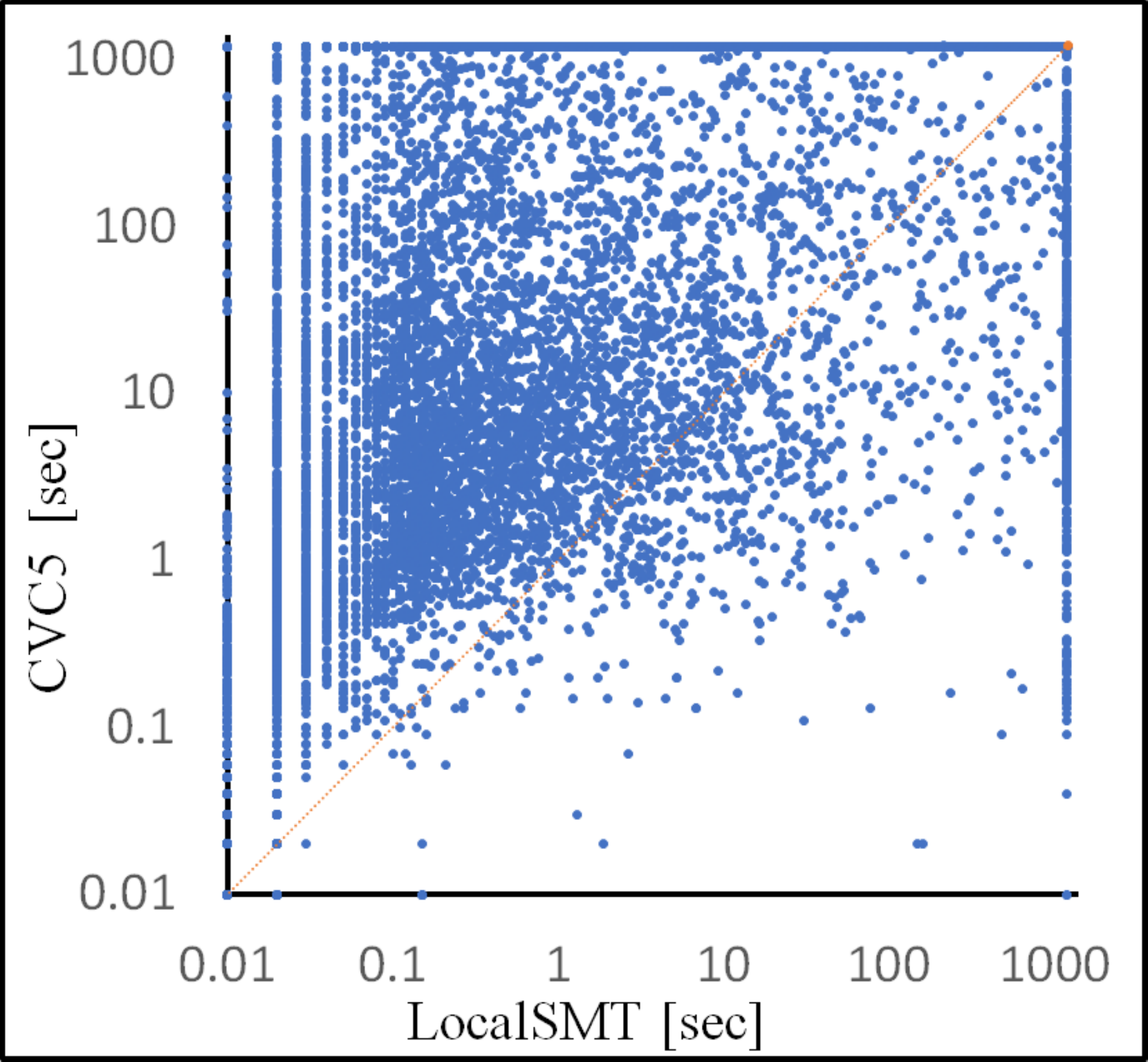}
         \caption{ Comparing with CVC5}
     \end{subfigure}

     \begin{subfigure}[b]{0.45\textwidth}
         \centering
         \includegraphics[width=5cm]{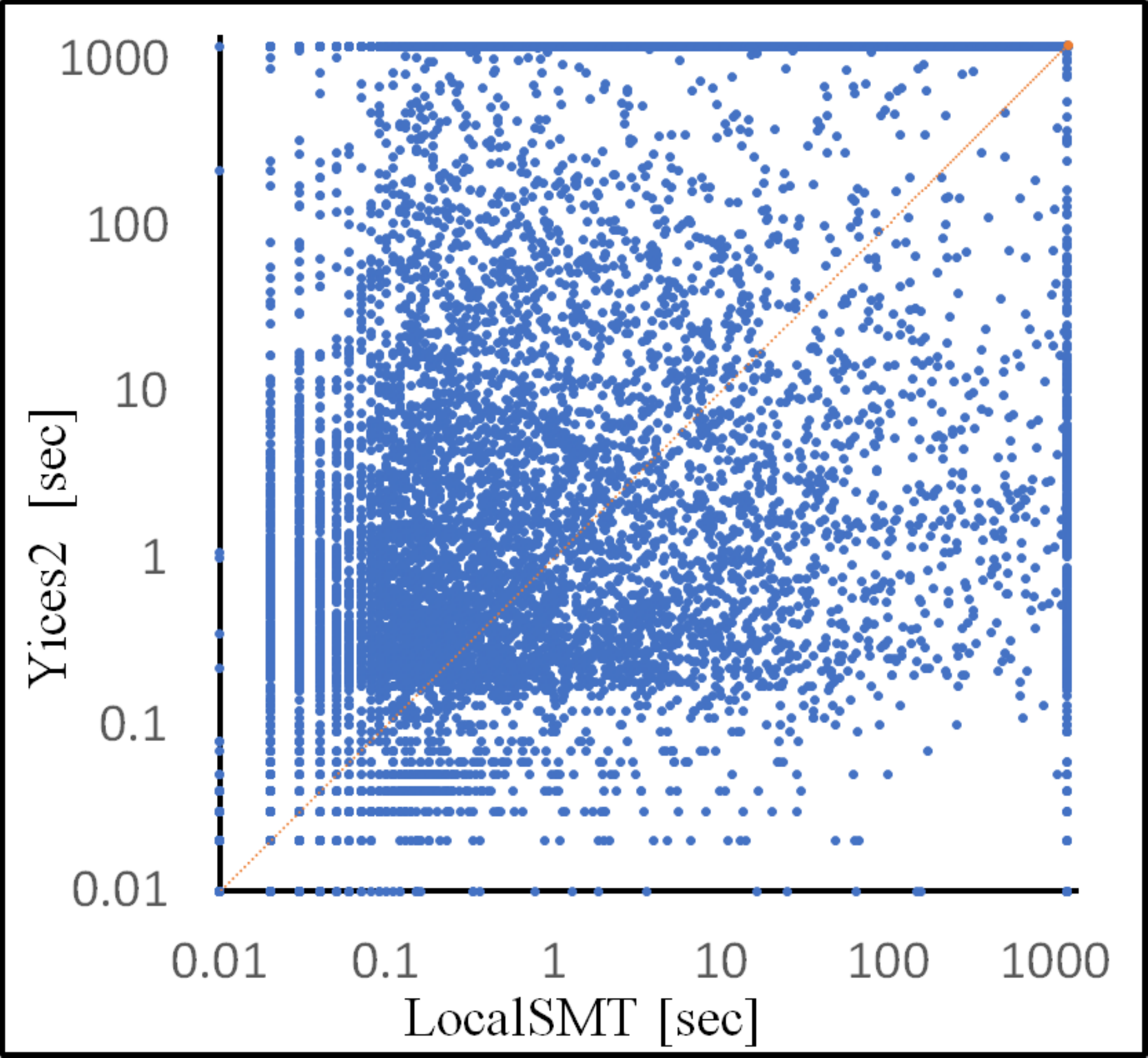}
         \caption{ Comparing with Yices2}
     \end{subfigure}
     \begin{subfigure}[b]{0.45\textwidth}
         \centering
         \includegraphics[width=5cm]{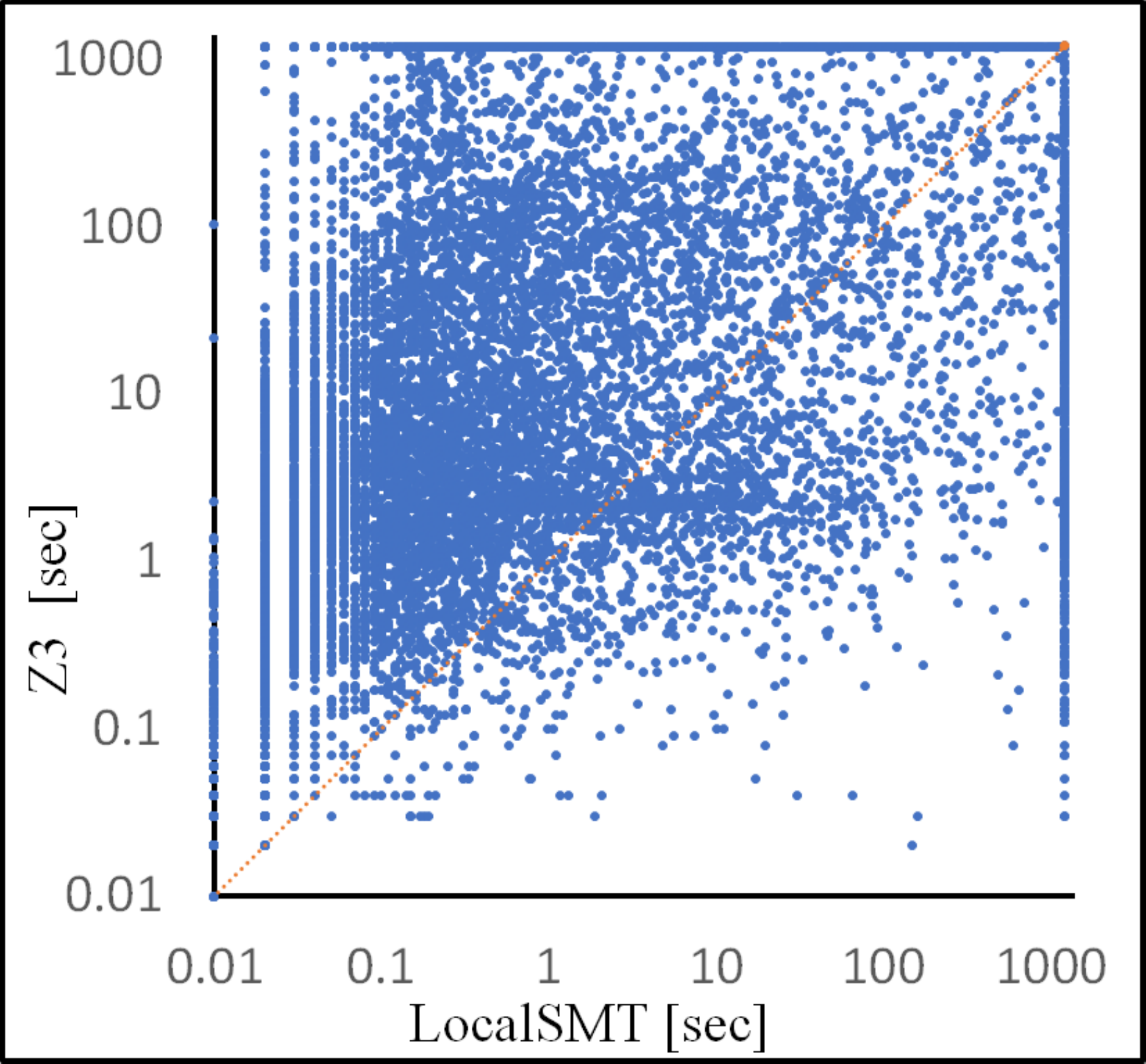}
         \caption{ Comparing with Z3}
     \end{subfigure}
        \caption{Run time comparison on Without Boolean category of SMTLIB-NIA}
        \label{fig:nia}
\end{figure}

\begin{figure}[t]
\centering
\centerline{\includegraphics[scale=0.6]{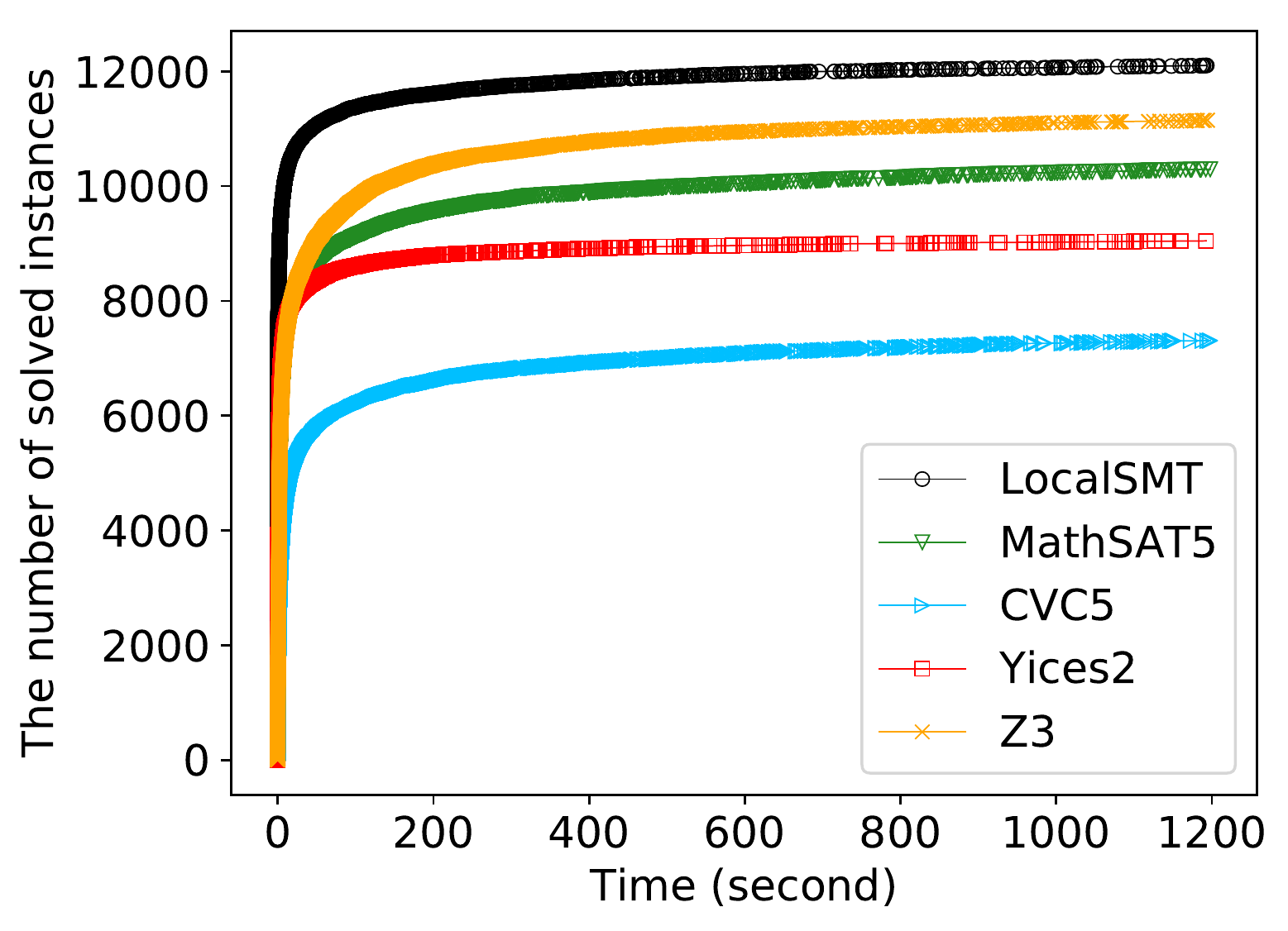}}
\caption{Run time distribution comparison with competitor on Without Boolean category of SMTLIB-NIA}
\label{distribution_nia_competitors}
\end{figure}

\subsection {\bf Results on SMTLIB-NIA Benchmark set (Table \ref{tbl:nia}, Fig. \ref{fig:nia})}
Our LocalSMT solver shows overwhelming performance on SMT(NIA) instances without Boolean variables (solving 12132 out of the 16439 instances), which can be seen from Table \ref{tbl:nia} and Fig. \ref{fig:nia}.
Note that there exist too many instances, making the run time comparison on Without Boolean category in Fig. \ref{fig:nia} indistinct, and thus we present the run time distribution comparison with competitors in Fig. \ref{distribution_nia_competitors}.
It works particularly well on ``ITS'' and ``CInteger'' instances, which are industrial instances for termination proving ~\cite{larraz2014proving}.
Moreover, LocalSMT can exclusively solve 75 ``mcm'' instances where no competitors can solve, which account for 41.4\% of such type of instances.

LocalSMT performs extremely poorly on instances of ``MathProblems(2022)'' type, failing to solve any benchmark of this type, for the following reasons:
First, the benchmark set contains hard mathematical problems such as  {\it sum of three cubes}~\cite{wooley2000sums} and {\it Magic squares of cubes}~\cite{andrews1917magic}, where the exponents on each variable is higher than 2. In this case, our solver cannot find any {\it critical move} operation as we only apply {\it critical move} when the exponent is not higher than 2, and thus it fails to find any solution.
Moreover, these benchmarks have few clauses, which are not typical SMT formulas, and hence the local-search scoring function concerning the state of clauses will become ineffective.

On those SMT(NIA) instances with Boolean variables, LocalSMT performs overall worse than its competitors, but is still comparable.
In general, LocalSMT can solve the most instances on SMTLIB-NIA.


\begin{table}[t]
\centering
\caption{{Summary results on SMTLIB-LIA, SMTLIB-IDL and SMTLIB-NIA.
Instances without and with Boolean variables are denoted by ``no\_bool'' and ``with\_bool'' respectively.}
}
\label{portfolio_table}
\setlength{\tabcolsep}{2.6mm}\scalebox{1}{
\begin{tabular}{@{}lllllll|l@{}}
\toprule
                & \#inst & MathSAT5 & CVC5 & Yices2       & Z3            & LocalSMT        & \textbf{Z3+LS} \\ \midrule
LIA\_no\_bool   & 6670 & 6442 & 6242 & 5994 & 6385  & \textbf{6478} & 6536 \\
LIA\_with\_bool & 1842 & 1619 & 766  & \textbf{1662} & 1617  & 912  & 1625 \\
Total           & 8512 & \textbf{8061} & 7008 & 7656 & 8002  & 7390 & \underline{8161} \\ \midrule
IDL\_no\_bool  & 841  & 363 & 539  & 654  & 653           & \textbf{687}  & 687   \\
IDL\_with\_bool & 770    & 514      & 586  & 658      & \textbf{665}  & 319           & 661   \\
Total           & 1611 & 877 & 1125 & 1312 & \textbf{1318}  & 1006           & \underline{1348}  \\ \bottomrule
NIA\_without\_bool & 16439  & 10497       & 7535        & 9157        & 11806           & \textbf{12132} & 12946 \\
NIA\_with\_bool     & 1980   & 1906        & 1908        & 1942       & \textbf{1959}    & 1669          & 1952 \\
Total              & 18419 & 12403         & 9443        & 11099      & 13765            & \textbf{13801} & \underline{14898} \\ \bottomrule
\end{tabular}
}
\end{table}

\begin{figure}[t]
\centering
\centerline{\includegraphics[scale=0.35]{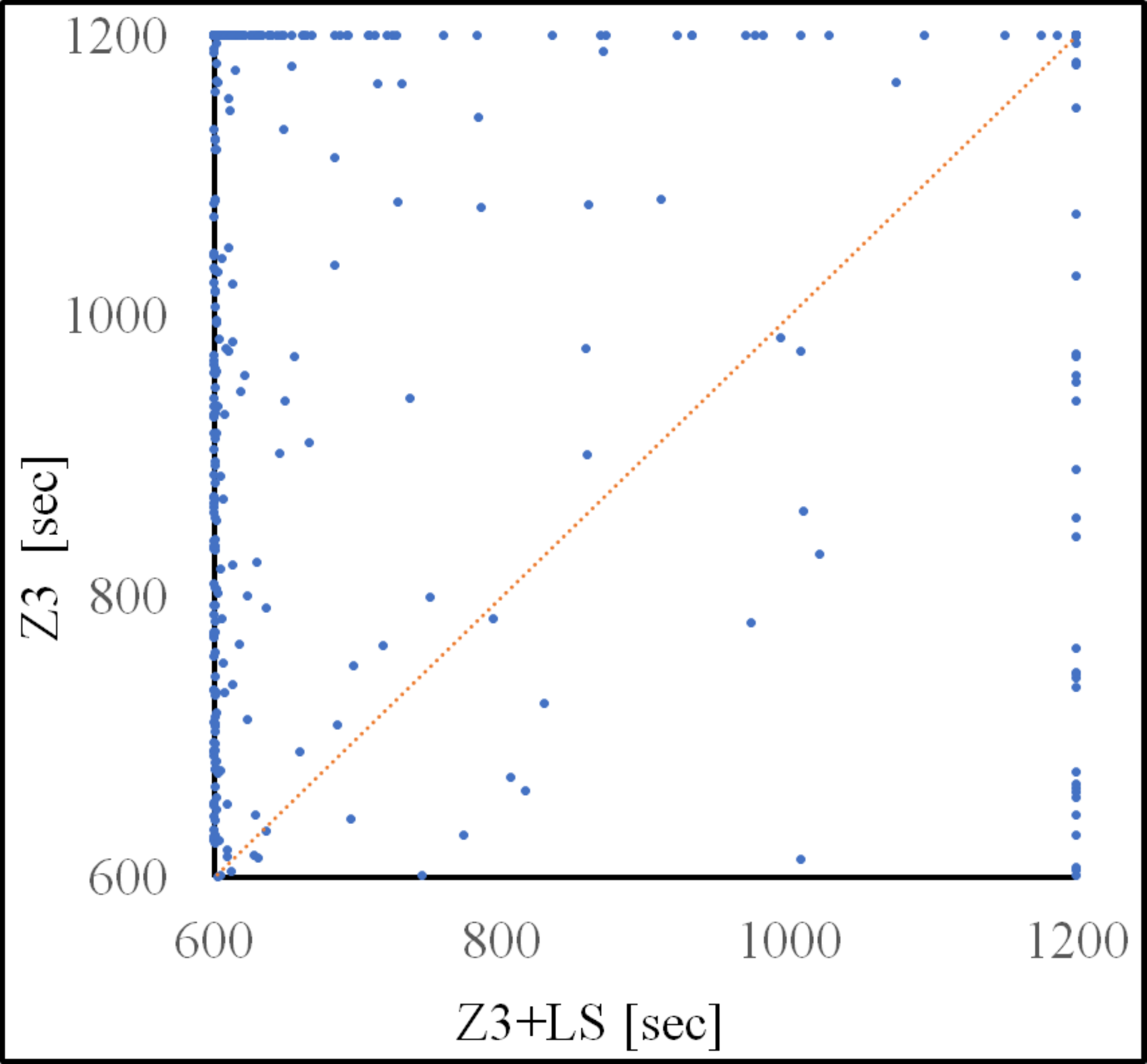}}
\caption{Run time comparison between Z3+LS and default Z3 on all benchmarks.}
\label{z3+ls}
\end{figure}

\subsection{\bf Combination with Z3 and Summary on SMT-LIB benchmark sets (Table \ref{portfolio_table})}
To confirm the complementarity of our local search solver with state of the art SMT solvers, we combine LocalSMT with Z3, by running Z3 with a time limit of 600s, and then LocalSMT from scratch with the remaining 600s if Z3 fails to solve the instance.
This wrapped solver can be regarded as a sequential portfolio solver, denoted as ``Z3+LS''.

We summarize the results of all solvers, including Z3+LS, on 3 SMT-LIB benchmark sets in Table \ref{portfolio_table}. 
Among all single-engine solvers, MathSAT5 solves the most instances of SMTLIB-LIA benchmark set, Z3 solves the most instances of SMTLIB-IDL benchmark set, while LocalSMT solves the most instances of SMTLIB-NIA benchmark set.
Moreover, LocalSMT outperforms its competitors on instances without Boolean variables from all 3 benchmark sets, indicating that local search is an effective approach for solving SMT(IA) instances with only integer variables.

Z3+LS solves more instances than any other solver on all benchmark sets, confirming that LocalSMT and Z3 have complementary performance and their  combination pushes the state of the art in solving satisfiable instances of SMT(IA). We also combined LocalSMT with Yices in the same manner, resulting in a wrapped solver called YicesLS ~\cite{caiyicesls}, which won the Single-Query and Model-Validation Track on QF\_IDL in SMT-COMP 2021.

The run time comparison between Z3+LS and the default Z3 on all benchmarks is presented in Fig. \ref{z3+ls}.
Since the portfolio solver Z3+LS runs Z3 for the first 600s, Z3+LS and Z3 show the same performance in the first 600s, and thus we only compare two solvers on the remaining 600s.
As shown in Fig \ref{z3+ls}, most points are concentrated on the top left side of the figure, indicating most instances which fails to be solved by Z3 can be solved fast once switching to running LocalSMT.
Hence, this confirms that LocalSMT is complementary to state-of-the-art SMT solvers.

Note that in contrast to the delicate portfolio configuration of previous work where the cutoff time for the local search approach is parameterized over metrics depending on the input problem~\cite{niemetz2016precise,niemetz2020ternary}, our portfolio configuration is set as basic as possible, 
because the main purpose of the portfolio configuration in this evaluation is to show that LocalSMT is indeed complimentary.
We want to demonstrate that even though the portfolio configuration may not be the optimal setting, Z3+LS still outperforms all rivals, and it shows that LocalSMT is able to solve problems that can be considered hard for Z3 alone, confirming the complementarity of LocalSMT with state-of-the-art SMT solvers.
Thus, we adopt a straightforward portfolio configuration, allocating equal time to both solvers.
Since neither solvers interacts with the other, the order does not influence the number of solutions; hence, it does not matter whether to Z3 or LocalSMT is  executed first.

\begin{table}[]
\caption{Comparison with incomplete solvers on SMTLIB-NIA}
\label{incomplete nia}
\begin{tabular}{@{}lllllllll@{}}
\toprule
Family                           & Type                & \#inst & bcl-maxsmt    & bcl-ninc-cores & bcl-ninc      & bcl-omt       & bcl-cores     & LocalSMT          \\ \midrule
\multirow{8}{*}{Without Boolean} & AProVE              & 1676   & 1662          & \textbf{1663}  & \textbf{1663} & 1648          & \textbf{1663} & 1627           \\
                                 & CInteger            & 1134   & 805           & 847            & \textbf{850}  & 818           & 849  & 801            \\
                                 & ITS                 & 12414  & 9216          & 9391           & 9402          & 9136          & 9397          & \textbf{9449}  \\
                                 & LassoRanker         & 4      & 4             & 4              & 4             & 4             & 4             & 4              \\
                                 & leipzig             & 162    & 153           & 155            & 155           & 147           & \textbf{158}  & 154            \\
                                 & mcm                 & 181    & 17            & 18             & 23            & 19            & 15            & \textbf{97}    \\
                                 & MathProblems(2022)        & 868    & 587            & 586             & 586            & \textbf{655}            & 646            & 0    \\
                                 &                     &        &               &                &               &               &               &                \\
                                 & Total               & 16439  & 12444         & 12664          & 12683         & 12427         & \textbf{12732}         & 12132 \\ \midrule
\multirow{7}{*}{With Boolean}    & calypto             & 80     & \textbf{80}   & \textbf{80}    & \textbf{80}   & \textbf{80}   & \textbf{80}   & 32             \\
                                 & Dartagnan(2021)  & 33     & 0             & 0              & 0             & 0             & 0             & 0              \\
                                 & ezsmt(2019)          & 8      & 0             & 0              & 0             & 0             & 0             & 0              \\
                                 & SAT14               & 1853   & \textbf{1853} & \textbf{1853}  & \textbf{1853} & \textbf{1853} & \textbf{1853} & 1633           \\
                                 & UltimateLassoRanker & 6      & \textbf{6}    & \textbf{6}     & \textbf{6}    & \textbf{6}    & \textbf{6}    & 4              \\
                                 &                     &        &               &                &               &               &               &                \\
                                 & Total               & 1980   & \textbf{1939} & \textbf{1939}  & \textbf{1939} & \textbf{1939} & \textbf{1939} & 1669           \\ \bottomrule
\end{tabular}
\end{table}

\subsection{Comparison with Incomplete solvers}
We compare LocalSMT with 5 modified variants of the incomplete solver proposed in ~\cite{borralleras2019incomplete}, where the SMT(NIA) problems are solved by reducing them to SMT(LIA).
Note that instances from CInteger and ITS benchmark sets are generated by their constraint-based termination prover VeryMax~\cite{borralleras2017proving} on the divisions of the termination competition termCOMP2016\footnote{https://termination-portal.org/wiki/Termination\_and\_Complexity\_Competition\_2016}, taking up a large proportion of SMTLIB-NIA, amount to 77.2\%.

The experimental results are shown in Table \ref{incomplete nia}.
LocalSMT is competitive but in general worse than the incomplete solvers, but it works particularly well on ``mcm'' type.
The key weakness compared with the competitors lies in the ``MathProblems(2022)'' type, which has been proved above to be unsuitable for LocalSMT.
Excluding the ``MathProblems(2022)'' type, LocalSMT can outperform all incomplete solver on the without Boolean category, which convinces again that LocalSMT performs particularly well on instances with only integer variables.

Note that there seems to be some issues in the incomplete solvers proposed in ~\cite{borralleras2017proving}. 
On 5 instances\footnote{These instances are as follows:

ConcurrencySafety-Main/28-race\_reach\_45-escape\_racing-O0.smt2,

ReachSafety-Loops/hard2\_unwindbound1-O0.smt2,

ReachSafety-Loops/hard2\_unwindbound2-O0.smt2,

ReachSafety-Loops/id\_trans-O0.smt2,

ReachSafety-Loops/verisec\_NetBSD-libc\_loop-O0.smt2.
} from ``Dartagnan'', all modified variants of the incomplete solver report ``UNSAT'', while 4 CDCL(T) competitors report ``SAT''.

\subsection{ Effectiveness of Proposed Strategies}
To analyze the effectiveness of the strategies in LocalSMT, we modify  LocalSMT to obtain 5 alternative versions as follows.
\begin{itemize}
    
    \item To analyze the effectiveness of the {\it cm} operator, we modify LocalSMT by replacing the {\it cm} operator with the operator that directly modifies an integer variable by a fixed increment $inc$, leading to two versions v\_fix\_1 and v\_fix\_5, where $inc$ is set as 1 and 5 respectively.
    \item To analyze the effectiveness of the two level heuristic for picking a decreasing {\it cm} operation, we modify LocalSMT by choosing a decreasing {\it cm} operation only from falsified clauses or directly from all false literals, leading to two versions, namely v\_focused and v\_extend.
    \item To analyze the effectiveness of $dscore$, we modify LocalSMT to choose a {\it cm} operation with the highest $score$ from the selected clause at local optima, leading to the version v\_score.
 
\end{itemize}

\begin{figure}[t]
\centering
\centerline{\includegraphics[width=10cm]{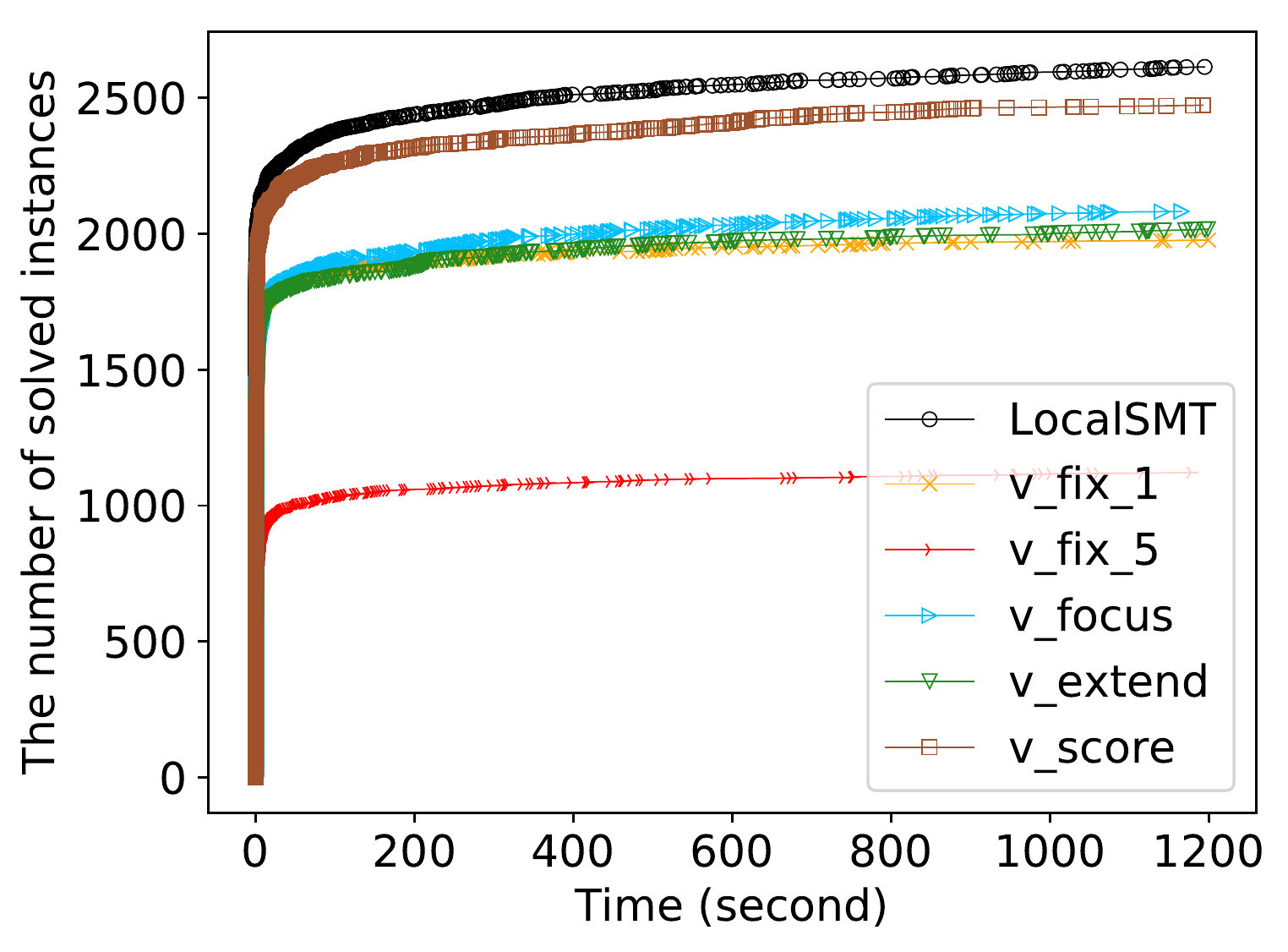}}
\caption{Run time distribution comparison with modified version on SMTLIB-LIA}
\label{distribution_lia}
\end{figure}

\begin{figure}[t]
\centering
\centerline{\includegraphics[width=10cm]{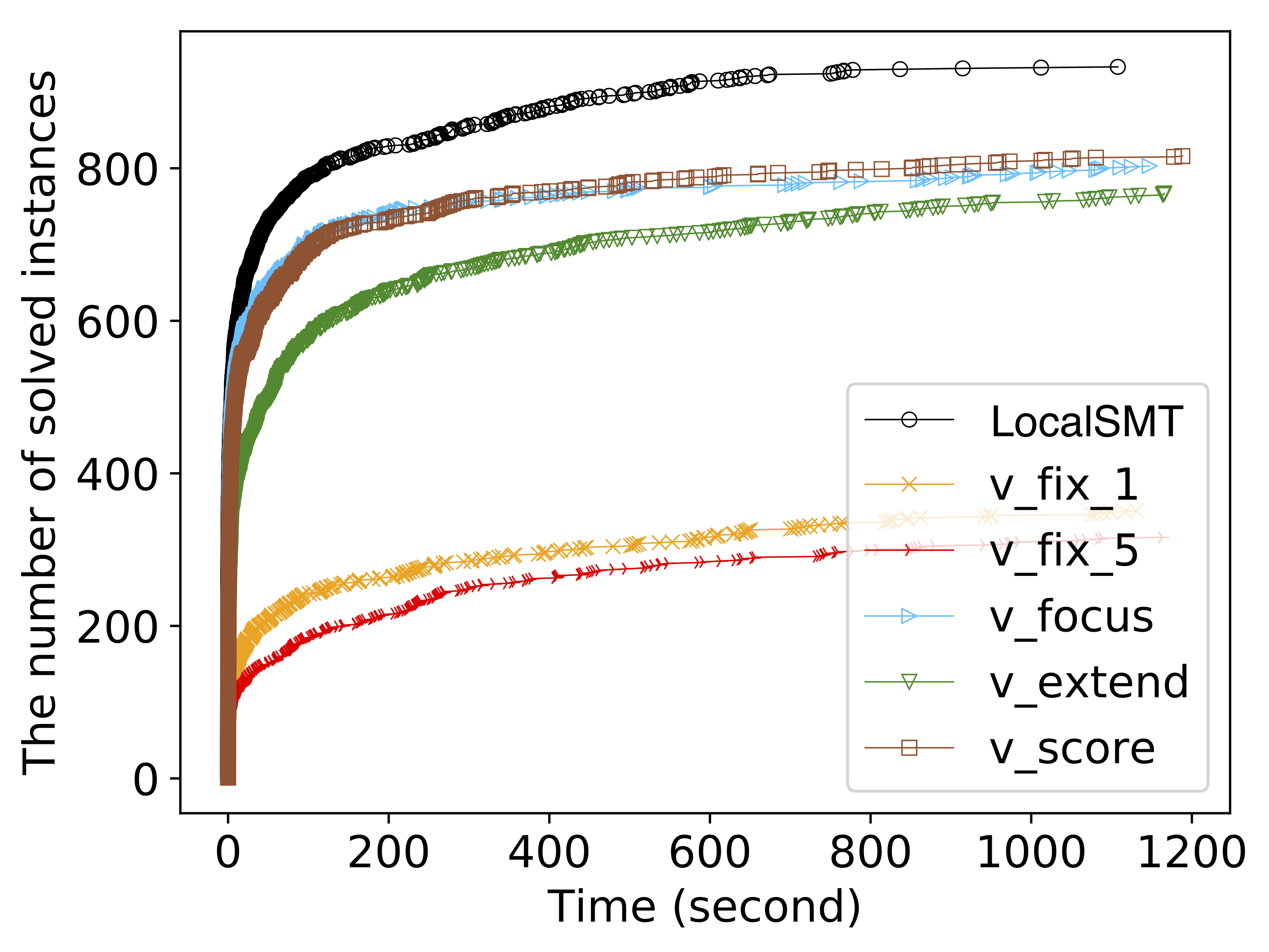}}
\caption{Run time distribution comparison with modified versions on SMTLIB-IDL}
\label{distribution_idl}
\end{figure}

\begin{figure}[t]
\centering
\centerline{\includegraphics[width=10cm]{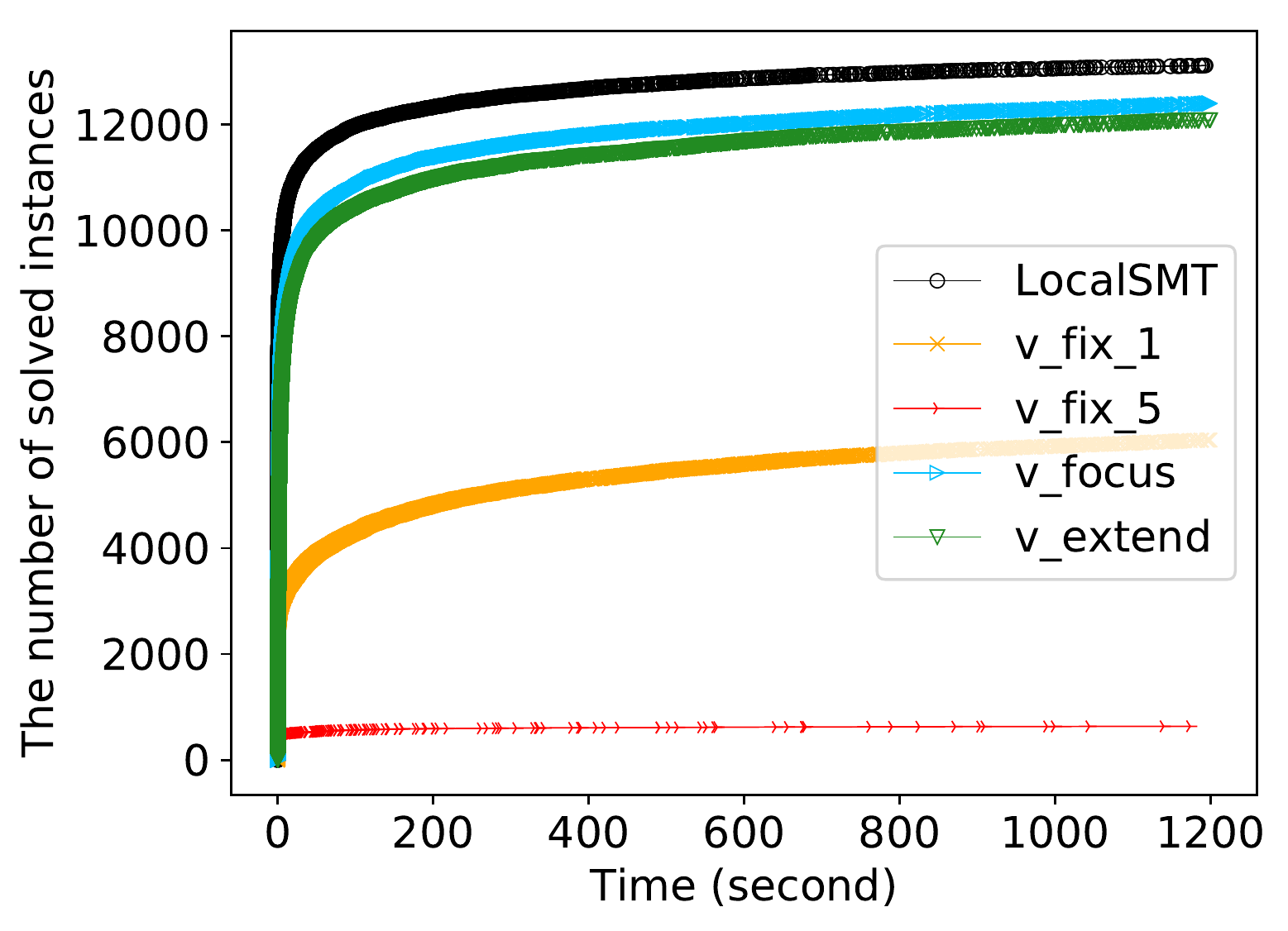}}
\caption{Run time distribution comparison with modified versions on SMTLIB-NIA}
\label{distribution_nia}
\end{figure}

We compare LocalSMT with these modified versions on the SMTLIB-LIA and SMTLIB-IDL benchmark sets.
The runtime distribution of LocalSMT and its modified versions on the two benchmark sets are presented in Figure \ref{distribution_lia}-\ref{distribution_nia}, confirming the effectiveness of the strategies.
Note that instances which can be solved by all versions within 1 second are dismissed, specifically, such instances in SMTLIB-LIA, SMTLIB-IDL and SMTLIB-NIA benchmark sets amount to 4759, 73 and 505 respectively.

From the runtime distribution comparison, we can conclude that {\it critical move} is the most important component to improve performance, because LocalSMT can significantly outperform  v\_fix\_1 and v\_fix\_5.
A possible explanation for the great promotion effect of {\it critical move} is that the search space of SMT(IA) is sparse, and thus modifying the variable to threshold can help the algorithm avoid getting stuck in a space without solution.
The idea of two-level heuristic also plays an important role in our algorithm. Comparatively, the {\it dscore} is less effective, although it also contributes considerably to the SMTLIB-IDL instances. One explanation is that the calculation of {\it dscore} requires much overhead, which reduces the benefits it brings.

\section{Conclusion and Future Work}
We developed the first local search solver for SMT(IA), opening the local search direction for SMT on integer theories.
 Main features of our solver include a  framework switching between Boolean and Integer modes,  the critical move operator and a scoring function based on distance to satisfaction.
Experiments show that our solver is competitive and complementary to  state-of-the-art SMT solvers.

We would like to enhance our solver by improving the performance on instances with Boolean variables, and develop local search solver for SMT on real number theories.
It is also interesting to explore deep cooperation with DPLL(T)  solvers.

\section*{Acknowledgements}{This work is supported by the Strategic Priority Research Program of the Chinese Academy of Sciences, Grant No. XDA0320000 and XDA0320300, and NSFC Grant 62122078.}

\bibliographystyle{ACM-Reference-Format}
\bibliography{sample-base}


\end{document}